\documentclass[sigconf]{acmart}
\AtBeginDocument{%
  }

\copyrightyear{2026}
\acmYear{2026}
\setcopyright{cc}
\setcctype{by}
\acmConference[WWW '26]{Proceedings of the ACM Web Conference 2026}{April 13--17, 2026}{Dubai, United Arab Emirates}
\acmBooktitle{Proceedings of the ACM Web Conference 2026 (WWW '26), April 13--17, 2026, Dubai, United Arab Emirates}
\acmPrice{}
\acmDOI{10.1145/3774904.3792228}
\acmISBN{979-8-4007-2307-0/2026/04}
\usepackage{amsmath}
\usepackage{graphicx}
\usepackage{enumitem}
\usepackage{amsthm}
\usepackage{multirow}
\usepackage{subcaption}
\usepackage{xspace}
\usepackage{booktabs}
\usepackage{lipsum} 
\usepackage{algorithm, algorithmic}
\usepackage{threeparttable} 
\usepackage{balance}
\usepackage{pifont}

\usepackage{makecell}
\usepackage{placeins}

\newcommand{\SR}[0]{SocialRec\xspace}

\newcommand{\modulecommunity}[0]{CEG\xspace}
\newcommand{\modulecomlong}[0]{\emph{community-aware user embedding generation}\xspace}
\newcommand{\moduleindividual}[0]{SIEG\xspace}
\newcommand{\moduleindlong}[0]{\emph{socially-connected item-aware user embedding generation}\xspace}
\newcommand{\method}[0]{\textsc{PULSE}\xspace}

\newcommand{\methodlong}[0]{\underline{P}arameter-efficient \underline{U}ser representation \underline{L}earning with \underline{S}ocial Knowledg\underline{e}\xspace}

\begin{document}

\title{PULSE: Socially-Aware User Representation Modeling Toward Parameter-Efficient Graph Collaborative Filtering}
\author{Doyun Choi}
\authornote{Both authors contributed equally to this research.}
\affiliation{%
  \institution{Korea Advanced Institute of Science and Technology}
  \city{Daejeon}
  \country{South Korea}}
\email{doyun.choi@kaist.ac.kr}

\author{Cheonwoo Lee}
\authornotemark[1]
\affiliation{%
  \institution{Korea Advanced Institute of Science and Technology}
  \city{Daejeon}
  \country{South Korea}}
\email{cheonwoo.lee@kaist.ac.kr}

\author{Biniyam Aschalew Tolera}
\affiliation{%
  \institution{Korea Advanced Institute of Science and Technology}
  \city{Daejeon}
  \country{South Korea}}
\email{binasc@kaist.ac.kr}

\author{Taewook Ham}
\affiliation{%
  \institution{Korea Advanced Institute of Science and Technology}
  \city{Daejeon}
  \country{South Korea}}
\email{htw0618@kaist.ac.kr}

\author{Chanyoung Park}
\affiliation{%
  \institution{Korea Advanced Institute of Science and Technology}
  \city{Daejeon}
  \country{South Korea}}
\email{cy.park@kaist.ac.kr}

\author{Jaemin Yoo}
\affiliation{%
  \institution{Korea Advanced Institute of Science and Technology}
  \city{Daejeon}
  \country{South Korea}}
\email{jaemin@kaist.ac.kr}

\renewcommand{\shortauthors}{Doyun Choi et al.}
\begin{abstract}
Graph-based social recommendation (\SR) has emerged as a powerful extension of graph collaborative filtering (GCF), which leverages graph neural networks (GNNs) to capture multi-hop collaborative signals from user-item interactions.
These methods enrich user representations by incorporating social network information into GCF, thereby integrating additional collaborative signals from social relations. However, existing GCF and graph-based \SR approaches face significant challenges: they incur high computational costs and suffer from limited scalability due to the large number of parameters required to assign explicit embeddings to all users and items.
In this work, we propose \method (\methodlong), a framework that addresses this limitation by constructing user representations from socially meaningful signals without creating an explicit learnable embedding for each user.
\method reduces the parameter size by up to 50\% compared to the most lightweight GCF baseline. Beyond parameter efficiency, our method achieves state-of-the-art performance, outperforming 13 GCF and graph-based social recommendation baselines across varying levels of interaction sparsity, from cold-start to highly active users, through a time- and memory-efficient modeling process.
Our implementation is available at \url{https://github.com/cdy9777/PULSE}.

\end{abstract}



\maketitle
\newcommand\webconfavailabilityurl{https://doi.org/10.5281/zenodo.18295272}
\ifdefempty{\webconfavailabilityurl}{}{
\begingroup\small\noindent\raggedright\textbf{Resource Availability:}\\
The source code of this paper has been made publicly available at \url{\webconfavailabilityurl}.
\endgroup
}

\section{Introduction}

Recommender systems have become indispensable tools for providing personalized experiences across various domains, such as e-commerce \cite{schafer1999recommender, linden2003amazon}, online streaming \cite{gomez2015netflix, harper2015movielens}, and social networking platforms \cite{abel2011analyzing, abel2013twitter}.
Traditional systems rely on interaction histories between users and items to model user preferences.
However, in real-world scenarios, most interactions are observed implicitly through behaviors such as clicks or views \cite{hu2008collaborative}. This implicit setting introduces ambiguity, making it difficult to reliably capture user intent. Social recommendation (\SR) \cite{10.1145/1458082.1458205, Guo_Zhang_Yorke-Smith_2015} has emerged as a promising alternative by incorporating social relationships between users as auxiliary signals. It utilizes the social network for more accurate modeling of user behavior, based on social theories such as social homophily~\cite{NEURIPS2024_c9e20f70,annurev:/content/journals/10.1146/annurev.soc.27.1.415} and social influence \cite{doi:10.1177/0049124193022001006, hogg2004enhancing}.

Evolving from early approaches primarily based on matrix factorization \cite{10.1145/1864708.1864736, 10.1145/1458082.1458205}, the field has advanced significantly with the advent of graph collaborative filtering (GCF) \cite{10.1145/3331184.3331267, 10.1145/3397271.3401063, choi2025simple}. GCF leverages the power of graph neural networks (GNNs) to detect collaborative signals in user-item interactions, which can be naturally represented as a bipartite graph. This enables user and item representations to incorporate higher-order neighbor information during the GCF encoding process. Similarly, social networks can also be represented as graphs, allowing GNN-based models to capture higher-order relationships among users, as well as between users and items \cite{10.1145/3331184.3331214, LIAO2022595}.
These graphs are fed into GCF models \cite{10.1145/3397271.3401063, 10.1145/3331184.3331267, chen2024multi,ma2024robust}, either independently or in combination with user-item interactions, to generate enriched user representations and yield improved recommendation.
We refer to this class of methods as graph-based \SR.

In both GCF and graph-based \SR approaches, each user is assigned a parameterized embedding that is updated during training, since explicit user information is often unavailable.
However, this architecture implies that the number of learnable parameters grows linearly with the number of users. Such growth becomes a major bottleneck in these approaches, leading to extremely high computational costs for datasets with large user sets and raising scalability concerns \cite{liu2021learnable, zhang2023sharklightweightmodelcompression}.
Furthermore, given the sparsity of interaction data, it is difficult to train all user embeddings effectively, which intensifies overfitting issues \cite{10.1145/3331184.3331267, cai2023lightgclsimpleeffectivegraph}. This limitation remains unresolved in existing graph-based \SR methods and, even worse, is often exacerbated by the introduction of additional parameters for social information refinement \cite{10.1145/3726302.3730013,10.1145/3637528.3671807, 10.1145/3442381.3449844}.

\begin{figure}
\centering
\includegraphics[width=\linewidth]{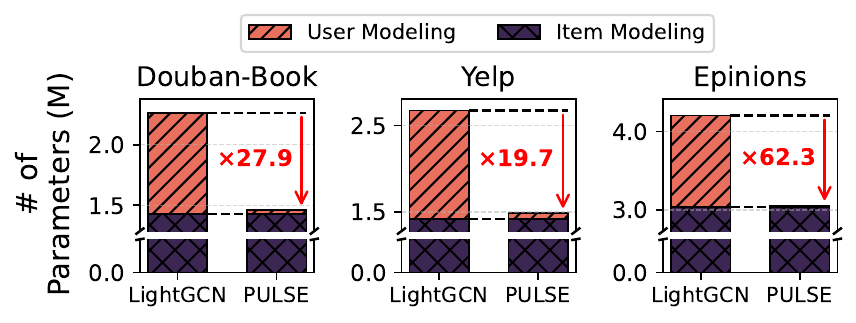}

\vspace{-10pt}
\caption{
    Comparison of the number of parameters between LightGCN and \method. \method reduces parameters by millions compared to LightGCN, the most lightweight GCF model.
}
\Description[Parameter reduction of PULSE compared to LightGCN]{
Parameter comparison between LightGCN and PULSE across three datasets.
The figure shows three grouped bar charts for Douban-Book, Yelp, and Epinions, reporting the number of model parameters in millions.
For each dataset, LightGCN consists of a large user modeling component and a smaller item modeling component, whereas PULSE relies primarily on item modeling with only a negligible user-related component.
As a result, the total number of parameters in PULSE is substantially smaller than that of LightGCN across all datasets.
The reduction ratios in user modeling parameters are explicitly annotated as 27.9 times on Douban-Book, 19.7 times on Yelp, and 62.3 times on Epinions, highlighting the near elimination of trainable user parameters in PULSE.
}
\vspace{-5pt}
\label{fig-parameter-analysis}
\end{figure}
In this work, we specifically highlight this limitation and propose \method (\methodlong), a new way for generating user representations in a highly parameter-efficient manner. Unlike prior graph-based \SR approaches, which focus on improving recommendation accuracy at the expense of additional computational costs, we aim to achieve both high accuracy and parameter-efficiency at the same time by leveraging social information as additional user features.



Specifically, from the perspective of user modeling, we use social information to construct socially aware user representations without relying on explicit learnable user embeddings.
We propose two modules to this end: \modulecomlong (\modulecommunity) and \moduleindlong (\moduleindividual). These modules incorporate different types of social signals into user modeling: community structures to which users belong and items with which their social neighbors have interacted. We further integrate these distinct signals in a user-adaptive manner and train all components through recommendation and self-supervised objectives to reduce potential noise that may occur from our novel user modeling architecture.
Figure~\ref{fig-parameter-analysis} shows that our approach saves the majority of parameters that would otherwise be required for assigning learnable embeddings to all users, reducing millions of parameters compared to the most lightweight GCF baseline.


We conduct extensive experiments to evaluate the effectiveness of \method across diverse recommendation scenarios. \method consistently outperforms 13 GCF and graph-based \SR baselines across three datasets, with clear performance margins. Moreover, it remains robust under varying levels of interaction sparsity, from cold-start to highly active users, as well as under noisy social networks. All these results are achieved with highly efficient computational cost. Our main contributions are summarized as follows:
\begin{itemize}[left=0cm] 

\item \textbf{Novelty:} To the best of our knowledge, \method is the first trial to propose a new type of socially-aware embedding generation as an alternative to conventional graph-based user modeling.
This is even achieved in a highly parameter-efficient manner.

\item \textbf{Efficiency:} \method significantly reduces the parameter size of graph-based recommendation by up to 50\% compared to LightGCN \cite{10.1145/3397271.3401063}.
It also exhibits linear time complexity with respect to the size of user-item interactions and the social network.

\item \textbf{Performance:} Our method achieves substantial performance gains in recommendation, with NDCG@20 increasing by 9.0\%, 8.1\%, and 8.5\% over the best competitors across three datasets. It also demonstrates robustness for users with sparse interactions as well as under noisy social network conditions.

\end{itemize}

\section{Problem and Related Works}
\label{sec:problem_and_related_works}


\subsection{Problem Definition}

Let $\mathcal{U}=\{u_k\}_{k=1}^m$ be the set of users and $\mathcal{I}=\{i_k\}_{k=1}^n$ be the set of items, where $m$ is the number of users and $n$ is the number of items.
Let $\mathbf{R} \in \{0, 1\}^{m \times n}$ be the interaction matrix between users and items, where $r_{u,i}=1$ if user $u$ has interacted with item $i$ and $r_{u,i} = 0$ otherwise.
In addition, let $\mathbf{S} \in \{0, 1\}^{m \times m}$ be an (undirected) social relation matrix between users, where $s_{u,v}=1$ if user $u$ is socially connected to user $v$ and $s_{u,v} = 0$ otherwise.
The goal of \SR is to predict user behavior toward items, i.e., recommendation, by utilizing the given $\mathbf{R}$ in conjunction with $\mathbf{S}$.

Graph-based \SR models the data as graphs to generate better representations of users and items.
For $\mathbf{R}$, we construct a bipartite graph $\mathcal{G}_R = (\mathcal{V}_R, \mathcal{E}_R)$, where the node set is $\mathcal{V}_R = \mathcal{U} \cup \mathcal{I}$ and the edge set is $\mathcal{E}_R = \{(u, i) \mid u \in \mathcal{U}, i \in \mathcal{I}, r_{u,i} = 1\}$. This graph can be represented by the following adjacency matrix:
\begin{equation}
    \label{eq:Adjacency matrix}
\mathbf{A}_R = \begin{bmatrix} 
\mathbf{0} & \mathbf{R} \\
\mathbf{R}^\top & \mathbf{0}
\end{bmatrix}.
\end{equation}
Similarly, we construct a social graph $\mathcal{G}_S = (\mathcal{V}_S, \mathcal{E}_S)$, with the node set $\mathcal{V}_S = \mathcal{U}$ and the edge set $\mathcal{E}_S = \{(u, v) \mid u,v \in \mathcal{U}, s_{u,v} = 1\}$.
The social relation matrix $\mathbf{S}$ itself serves as the adjacency matrix $\mathbf{A}_S$.

\subsection{Graph-Based Social Recommendation}
\SR methods have emerged as powerful tools for enhancing user modeling in recommender systems by integrating social trust information into user embeddings, enabling them to capture dependencies between users \cite{10.1145/1864708.1864736, Guo_Zhang_Yorke-Smith_2015}.
With the advent of graph collaborative filtering (GCF), which leverages graph neural networks (GNNs) to incorporate multi-hop collaborative signals from user-item interaction histories via message propagation, a variety of methods have been proposed to exploit social network structures as graphs and integrate them into recommender systems, such as DiffNet \cite{10.1145/3331184.3331214}, RecoGNN \cite{10.1145/3357384.3357924}, and SocialLGN \cite{LIAO2022595}. 

Recent work in graph-based \SR has incorporated the principles of social networks into user modeling.
For example, LSIR \cite{liu2024learning} leverages the theory of social homophily, where users form groups with similar preferences when modeling user representations. In contrast, CGCL \cite{hu2023celebrity} focuses on social influence, capturing how highly influential individuals shape the preferences of others. DISGCN \cite{li2022disentangled} combines both perspectives. 

However, real-world social networks are often incomplete and contain unreliable links.
Many approaches have increasingly focused on denoising techniques to improve the robustness of recommendations.
Preference-driven mechanisms  remove weak or noisy edges to enhance the social network structure \cite{10.1145/3589334.3645460, 10.1145/3637528.3671807, quan2023robust, hu2024hierarchical}, while diffusion-based approaches suppress noise by propagating signals in latent spaces to preserve meaningful relations \cite{10.1145/3637528.3671958, 10.1145/3627673.3679630}.
Alongside these, self-supervised learning (SSL) has emerged as a powerful tool for enhancing embedding quality by improving the discriminability of learned representations \cite{10.1145/3442381.3449844, wang2023denoisedselfaugmentedlearningsocial, 10.1145/3447548.3467340}.

However, these methods still follow the conventional GCF paradigm, relying on learnable embedding parameters for users and items to generate their representations. Even worse, the introduction of additional parameters for refining social network information further deteriorates storage efficiency and poses scalability challenges when applied to large-scale datasets. In this work, we propose a new way of leveraging social information that shifts the focus toward improving parameter efficiency and generalizability, while generating robust user representations well-suited 
for recommendation tasks.
\section{Proposed Method}
\label{sec:method}

We introduce \method, our framework for incorporating social information into user representations in a parameter-efficient manner;
our goal is to avoid having dedicated parameters for each individual user by leveraging social signals in a principled manner.

\begin{figure}
\centering
\includegraphics[width=\linewidth, trim={0.7cm 0 0.7cm 0},clip]{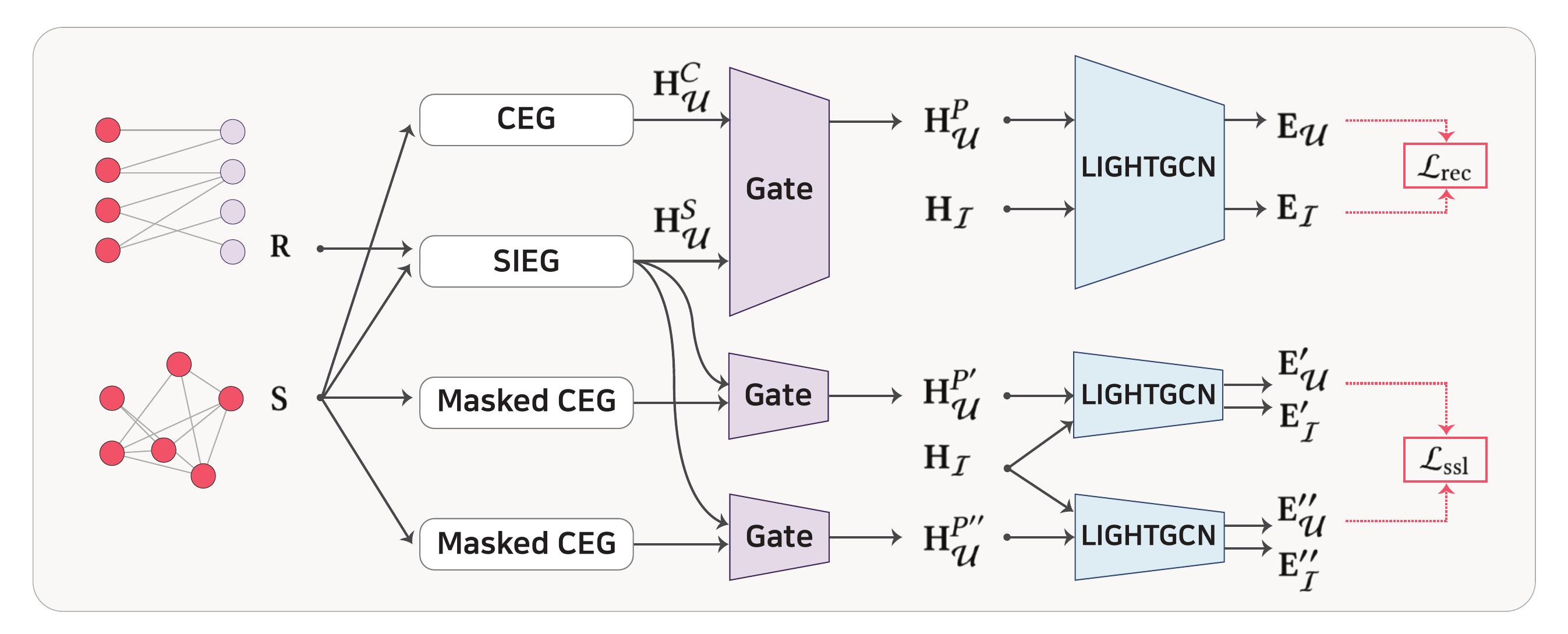}
\vspace{-20pt}
\caption{
    The overall framework of \method. \modulecommunity incorporates community information for each user, while \moduleindividual integrates item information from social neighbors. These distinct signals are fused adaptively for each user through a gating network. The resulting user embeddings are passed to LightGCN with item embeddings, and optimized using a recommendation and an additional self-supervised loss.
    Refer to Figure \ref{fig-moduel-detail} and \ref{fig-ssl-detail} for more detailed illustration of the components.
}
\Description[Overall architecture of PULSE with supervised and self-supervised branches]{
Overview of the PULSE framework for user representation learning.
The figure illustrates how interaction data and social information are processed through multiple graph encoders and gating modules.
The interaction graph and the social graph are first encoded by separate modules, including a community encoder and a social information encoder, producing multiple user representations.
These representations are adaptively fused by gate modules to generate personalized user embeddings.
The upper branch feeds the fused representations into a LightGCN model optimized with a recommendation loss, producing user and item embeddings.
The lower branches apply masked versions of the community encoder to construct multiple augmented views, which are passed through LightGCN to generate alternative embeddings.
A self-supervised learning loss is applied to align the embeddings from different augmented views, encouraging robust and consistent user representations.
}

\vspace{-15pt}
\label{fig-overall-framework}
\end{figure}

Figure \ref{fig-overall-framework} illustrates the overall structure of \method, which comprises three key components for incorporating disentangled social information into user representations:
(a) \modulecomlong (\modulecommunity), (b) \moduleindlong (\moduleindividual), and (c) an adaptive fusion module for disentangled social information. Beyond the architectural design, we further introduce a self-supervised learning objective that mitigates the impact of misleading biases in user modeling.

\subsection{Overview}
\label{ssec:overview}

At its core, \method generates user embeddings using \modulecommunity and \moduleindividual, which incorporate socially meaningful signals into user representations from distinct perspectives: higher-order community structures and items interacted with by social neighbors, respectively.
The outputs of these modules, $\mathbf{h}_u^{\text{C}}$ and $\mathbf{h}_u^{\text{S}}$, are combined to form the socially informed representation of each user $u$:
\begin{equation}
\label{eq:adaptiveagg}
\mathbf{h}_{u}^{\text{P}} = \alpha_u \mathbf{h}_u^{\text{C}} + (1-\alpha_u) \mathbf{h}_u^{\text{S}},
\end{equation}
where $\mathbf{h}_u^{\text{C}}$ and $\mathbf{h}_u^{\text{S}}$ denote the user representations from \modulecommunity and \moduleindividual, respectively. The parameter $\alpha_u$ adaptively balances the contributions of the two signals for user $u$. 
Details on $\alpha_u$ are provided in Section~\ref{ssec:aggregation}.

\paragraph{Encoding and Prediction}

After generating socially-aware user embeddings
$\mathbf{H}_{\mathcal{U}}^{\text{P}} \in \mathbb{R}^{m \times d}$, we provide them to a graph collaborative filtering (GCF) encoder, along with the learnable item embeddings $\mathbf{H}_\mathcal{I} \in \mathbb{R}^{n \times d}$ and the adjacency matrix $\mathbf{A}_R$:
\begin{equation}
\mathbf{E}_\mathcal{U},
\mathbf{E}_\mathcal{I}
= \text{GCF}([\mathbf{H}_{\mathcal{U}}^{\text{P}};\mathbf{H}_\mathcal{I}],\mathbf{A}_R).
\end{equation}

In this work, we adopt LightGCN \cite{10.1145/3397271.3401063} as the backbone encoder, which is parameter-efficient yet achieves strong recommendation performance.
It updates user and item embeddings as follows:
\begin{equation}
    \label{eq: LightGCN aggregation}
    \mathbf{E}^{(l)} = \mathbf{D}_R^{-1/2} \mathbf{A}_{R} \mathbf{D}_R^{-1/2}\mathbf{E}^{(l-1)}, \quad l = 1, 2, \dots, L,
\end{equation}
where $\mathbf{E}^{(l)} \in \mathbb{R}^{(m+n) \times d}$ contains both item and user embeddings at layer $l$, $\mathbf{D}_R$ is the degree matrix of $\mathbf{A}_{R}$, and the initial embeddings are $\mathbf{E}^{(0)} = [\mathbf{H}_{\mathcal{U}}^{\text{P}} ; \mathbf{H}_\mathcal{I}]$.
Then, the representations are merged as:
\begin{equation}
    \label{eq:final backbone representation}
    \textstyle \mathbf{E}_{\mathcal{U}} =  \sum_{l=0}^{L}\mathbf{E}_{\mathcal{U}}^{(l)}, \quad \mathbf{E}_{\mathcal{I}} = \sum_{l=0}^{L} \mathbf{E}_{\mathcal{I}}^{(l)}.
\end{equation}

The resulting $\mathbf{E}_{\mathcal{U}}$ and $\mathbf{E}_{\mathcal{I}}$ serve as the final representations of users and items for predictions.
The prediction for user $u$ and item $i$ is done by computing the dot product: $\hat{r}_{u,i} = \mathbf{e}_u^\top \mathbf{e}_i.$

\paragraph{Training}

The training of \method is done as multi-task learning that integrates the recommendation task with self-supervised regularization. The objective function is given as:
\begin{equation}
    \mathcal{L}(\theta) = \mathcal{L}_{\text{rec}}(\theta) + \lambda_{1} \mathcal{L}_{\text{ssl}}(\theta) + 
    \lambda_{2}||\theta||^{2}_{2},
\label{eq:objective-function}
\end{equation}
where $\theta$ is the set of all trainable parameters, $\mathcal{L}_{\text{rec}}$ is the loss for the recommendation task, $\mathcal{L}_{\text{ssl}}$ is the self-supervised loss, and the L2 regularization term is applied.
The coefficients $\lambda_1$ and $\lambda_2$ control the balance between the three terms during training.

We adopt the Bayesian Personalized Ranking (BPR) \cite{rendle2012bprbayesianpersonalizedranking} as the recommendation loss $\mathcal{L}_{\text{rec}}$ as follows:
\begin{equation}
    \mathcal{L}_{\text{rec}}(\theta) = - {\textstyle \sum_{(u,i,j) \in \mathcal{D}}} \log \sigma (\hat{r}_{u,i} - \hat{r}_{u,j}),
\end{equation}
where $\mathcal{D}=\{(u, i, j): r_{u,i}=1, r_{u,j}=0\}$ contains both positive and negative samples, and $\sigma$ is the sigmoid function.
$\mathcal{L}_{\text{rec}}$ encourages the model to assign higher predicted scores to positive interactions compared to negative ones, as done in typical GCF models.

The detailed formulation of the self-supervised loss $\mathcal{L}_{\text{ssl}}$ will be provided in Section \ref{ssec:ssl}, after introducing our main components.

\begin{figure*}[t]
    \centering
    \vspace{-20pt}
   \includegraphics[width=\textwidth,trim={0.5cm 0 0.1cm 0},clip]{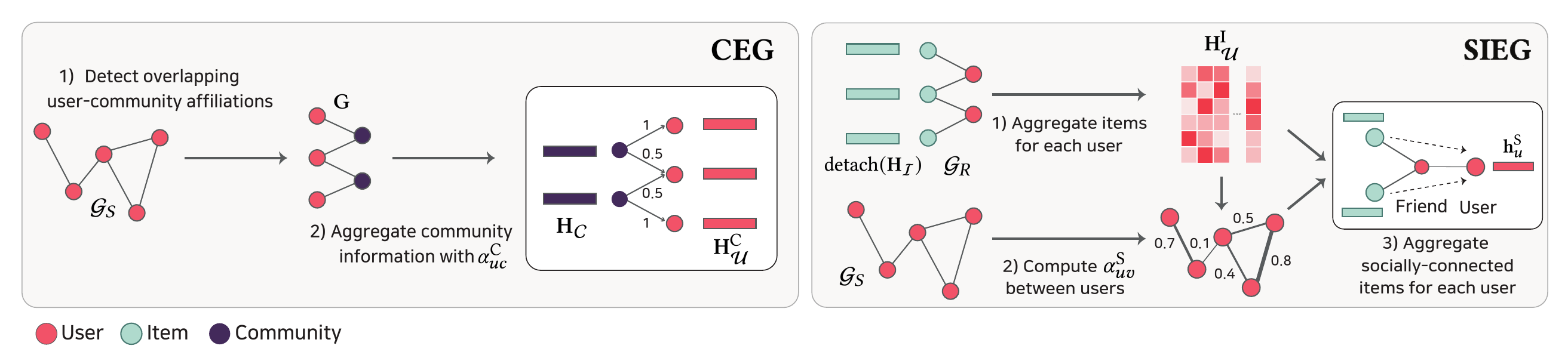}
    
    \vspace{-10pt}
    \caption{
        Visualization of the key modules. 
        (Left) \modulecommunity  leverages overlapping user-community affiliation information obtained through a community detection algorithm.
        It generates $\mathbf{H}_{\mathcal{U}}^{\text{C}}$ by aggregating community embeddings with weights $\alpha_{uv}^{\text{C}}$, determined by social user degrees in $\mathbf{G}$.
        (Right) \moduleindividual  incorporates socially-connected item information; it computes \smash{$\alpha_{uv}^{\text{S}}$} based on behavioral similarity between users, which is then applied in the aggregation of socially-connected items into $\mathbf{H}_{\mathcal{U}}^{\text{S}}$.
    }
    \vspace{-5pt}
    \Description[Illustration of community and socially connected item encoders in PULSE]{
Illustration of the community encoder and the socially connected item encoder used in PULSE.
The left part shows the community encoder, which first detects overlapping user–community affiliations from the social graph and then aggregates community-level information for each user.
Users can be associated with multiple communities, and the corresponding community representations are combined to form a community-aware user representation.
The right part shows the socially connected item encoder.
Item representations are first aggregated for each user, and pairwise social importance between users is computed based on their social connections.
These importance weights are then used to aggregate items from socially connected users, producing a socially informed user representation.
Together, the two encoders capture complementary community-level and socially connected item signals for user modeling.
}

    \label{fig-moduel-detail}
\end{figure*}
\subsection{Community-Aware User Embedding Generation (CEG)}
\label{ssec:ceg}

Community structures provide a high-level abstraction of social relationships that shape users' preferences \cite{lancichinetti2010characterizing}. Identifying these structures helps reveal patterns in collective user behavior, as users with similar interests tend to belong to the same community across various domains \cite{gasparetti2021community}. The role of community information in social recommendation has been extensively explored \cite{sahebi2011community, li2015overlapping, tang2016recommendation, liu2019social, guan2021community, ni2023community, ni2024graph}.
Building on this insight, we propose \modulecomlong (\modulecommunity) for integrating community signals into user modeling in a simple and efficient manner.
An overview of \modulecommunity is shown on the left side of Figure~\ref{fig-moduel-detail}.

\paragraph{Community Detection}

Our first stage is to detect user communities inherent in the structure of the social network $\mathcal{G}_S$, while we have two main points to consider.
First, we aim to detect overlapping communities, allowing each user to be associated with multiple social groups, as users often belong to multiple social groups with diverse semantics (e.g., friends, family, and colleagues).
Second, given the massive scale of social networks, we require community detection techniques that are both scalable and effective. 
Among various design choices for overlapping community detection, we adopt a lightweight two-step procedure inspired by a recent approach~\cite{hieu2024overlappingcommunitydetectionalgorithms}.
While end-to-end community detection methods (e.g., differentiable clustering) are viable alternatives, they typically rely on additional learnable embeddings and costly iterative updates, which conflict with our efficiency objective.

First, we apply the Leiden algorithm \cite{Traag_2019}, a method well known for its scalability and effectiveness in community detection, to obtain an initial non-overlapping partition. Second, we convert this partition into an overlapping affiliation by iteratively adding a user to neighboring communities whenever the inclusion of that user increases the modularity of those communities beyond a fixed threshold.
Details of this algorithm are provided in Appendix~\ref{appendix-slpa}.

Based on the user communities, we obtain a user-community affiliation matrix $\mathbf{G} \in \mathbb{R}^{m \times |\mathcal{C}|}$, where $g_{u,c} = 1$ indicates that user $u$ belongs to community $c \in \mathcal{C}$. 
Here, $m$ and $|\mathcal{C}|$ denote the number of users and detected communities, respectively. 
For those communities,
we assign learnable community embeddings $\mathbf{H}_{\mathcal{C}} \in \mathbb{R}^{|\mathcal{C}| \times d}$, which are optimized end-to-end during training. 
By learning them jointly with the recommendation objective, users within the same community effectively share community-level information.

\paragraph{User Representations}

To effectively incorporate community signals into each user, we adopt a message propagation mechanism widely used in GNNs \cite{kipf2016semi, xu2018powerful}. The user-community affiliation matrix $\mathbf{G}$ is interpreted as the adjacency matrix of a bipartite graph $\mathcal{G}_G = (\mathcal{V}_G, \mathcal{E}_G)$ between users and communities, where the node set is $\mathcal{V}_G = \mathcal{U} \cup \mathcal{C}$ and the edge set is $\mathcal{E}_G = \{(u, c) \mid u \in \mathcal{U}, c \in \mathcal{C}, g_{u,c} = 1\}$.
Using this graph, we generate user embeddings that incorporate their associated community signals as follows:
\begin{equation} \label{eq:user embedding generation_1}
    \textstyle \mathbf{h}_{u}^{\text{C}} = \sum_{c \in \mathcal{N}_G(u)} \alpha_{uc}^{\text{C}} \mathbf{h}_{c},
\end{equation}
where $\mathcal{N}_G(u)$ denotes the set of communities to which user $u$ belongs, and $\alpha_{uc}^{\text{C}}$ controls the influence of each community $c$ on user $u$.
For simplicity, we set $\alpha_{uc}^{\text{C}} = |\mathcal{N}_G(u)|^{-1}$ such that each community signal contributes equally to modeling user representations.


\subsection{Socially-Connected Item-Aware User Embedding Generation (\moduleindividual)}
\label{ssec:sia}

The core idea of \moduleindlong (\moduleindividual) is to enrich user representations by incorporating signals from items beyond those directly connected, using social connections as an auxiliary information source.
According to the theory of social influence, individuals’ interaction behaviors often affect one another, and the items they purchase tend to be similar \cite{doi:10.1177/0049124193022001006, annurev:/content/journals/10.1146/annurev.psych.55.090902.142015}. From this perspective, we propose that items interacted with by a user’s friends serve as valuable signals for modeling the target user’s preferences \cite{zhao2014leveraging}. We incorporate this information into the user modeling process alongside directly connected item information and denote these items as \emph{socially connected items}.
The right side of Figure \ref{fig-moduel-detail} provides an overview of \moduleindividual.

\paragraph{Social Item Aggregation}

We first aggregate item signals for each user based on the items with which they have interacted:
\begin{equation}
\label{eq: behavior embedding}
    \mathbf{h}_u^\text{I} = \sum_{i \in \mathcal{N}_R(u)} \frac{1}{\sqrt{d_R(u)}\sqrt{d_R(i)}} \mathrm{detach}(\mathbf{h}_{i}),
\end{equation}
where $\mathcal{N}_R(u)$ denotes the set of items that user $u$ has interacted with, and $d_R(\cdot)$ is the degree of a node in the interaction matrix $\mathbf{R}$.
The detached embedding $\mathrm{detach}(\mathbf{h}_i)$ prevents the gradient from flowing to the learnable parameter $\mathbf{h}_i$ during this process.
This is to ensure that the item embeddings are learned only through the GCF encoding process and are not affected by the social signals.


We then aggregate these signals using a message propagation mechanism in the social graph $\mathcal{G}_S$ as follows:
\begin{equation} \label{eq: SPIA result}
\mathbf{h}_u^{\text{S}} = \sum_{v \in \mathcal{N}_S(u)}\frac{\alpha_{uv}^{\text{S}}}{\sqrt{d_S(u)}\sqrt{d_S(v)}}  \mathbf{h}_{v}^{\text{I}},
\end{equation}
where $\mathcal{N}_S(u)$ denotes the set of neighbors of user $u$ in $\mathbf{S}$, $d_S(\cdot)$ is the degree of a user in the social network $\mathcal{G}_S$, and $\alpha_{uv}^{\text{S}}$ is the attention weight that controls the influence of each neighbor on user $u$.

\begin{figure*}
    \centering
    \vspace{-5pt}
   \includegraphics[width=\textwidth]{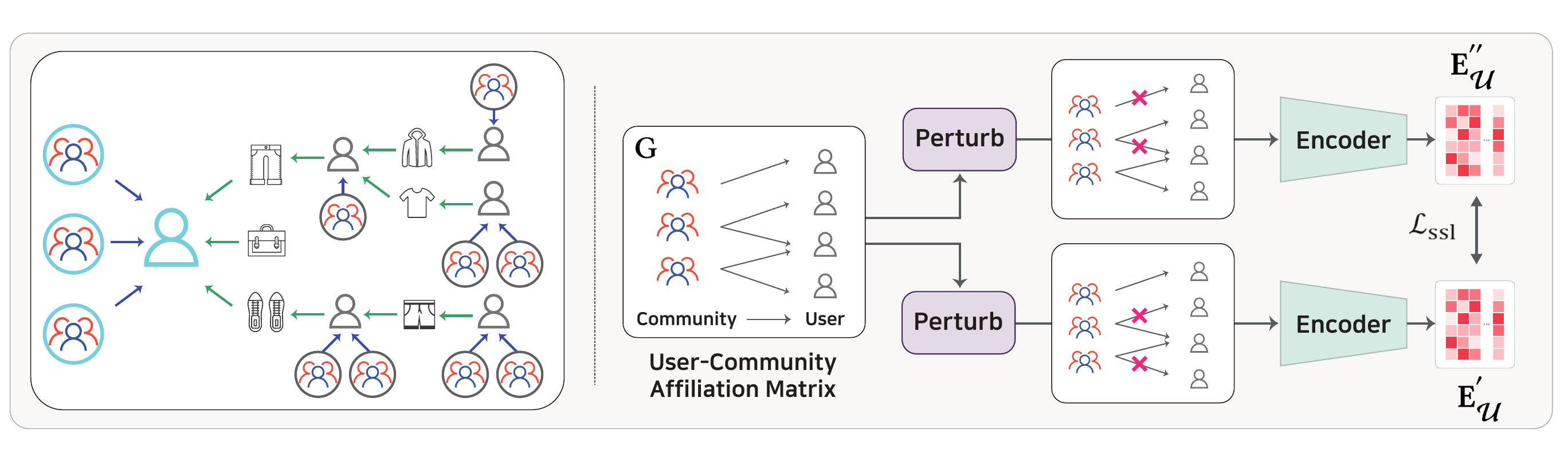}
    
    \vspace{-12pt}
    \caption{
        Illustration of self-supervised learning in \method.
        (Left) While generating community-aware user representations 
        $\mathbf{H}_\mathcal{U}^\text{C}$
        , the large receptive field from stacked GNN layers can introduce irrelevant community signals, potentially degrading the quality of representations. (Right) To address this, we create augmented views by randomly masking affiliation information in $\mathbf{G}$ and apply a self-supervised loss $\mathcal{L}_{\text{ssl}}$ that encourages the retained signals to align with the user’s most relevant communities.}
    \vspace{-5pt}
    \Description[Self-supervised augmentation via perturbation of user–community affiliations]{
Illustration of the self-supervised augmentation strategy based on user–community affiliations.
The left part illustrates which signals, including community-based and user–item interaction-based information, are incorporated into user representations during GNN-based encoding.
The user–community affiliation matrix is then perturbed to generate multiple augmented views by randomly removing a subset of user–community connections.
Each perturbed view is encoded independently to produce different user representations.
A self-supervised learning loss is applied to align the representations from the augmented views, encouraging robustness to variations in community affiliation structures.
}

    \label{fig-ssl-detail}

\end{figure*}

\paragraph{Social Attention}

As demonstrated by many existing SocialRec approaches \cite{10.1145/3589334.3645460, 10.1145/3627673.3679630}, social connections do not exert the same level of influence on the target user.
This highlights the need to account for user influence when aggregating item information.
We propose a mechanism to control the attention to aggregate item signals.

The user representations $\mathbf{H}_{\mathcal{U}}^{\text{I}}$ obtained by integrating item interaction signals represent users’ behavioral characteristics.
From this perspective, we compute the behavioral similarity for each socially connected pair of users $(u,v) \in \mathcal{E}_S$ as follows:
\begin{equation}
    \alpha^{\text{S}}_{uv} = \frac{1}{2}(1+\text{sim}(\mathbf{h}_u^{\text{I}}, \mathbf{h}_v^{\text{I}})) \exp \Bigl( -\frac{1}{2\sigma^2} \lVert {\mathbf{h}_u^{\text{I}}} - {\mathbf{h}_v^{\text{I}}} \rVert_2^2 \Bigr),
\end{equation}
where $\text{sim}(\mathbf{h}_u^{\text{I}}, \mathbf{h}_v^{\text{I}})$ denotes the cosine similarity between the embeddings of users $u$ and $v$.
This combines the normalized cosine similarity between ${\mathbf{h}_{u}^{\text{I}}}$ and ${\mathbf{h}_{v}^{\text{I}}}$ with a radial basis function (RBF) kernel, thereby capturing both angular alignment and magnitude consistency, as suggested by \cite{chen2024dual}.
We then normalize each similarity $\alpha_{uv}^{\text{S}}$ based on the degrees of users $u$ and $v$ in the social network as  
${\alpha_{uv}^{\text{S}}}/{\sqrt{d_S(u) d_S(v)}}$.
This mitigates the risk that high-degree nodes dominate representations by mixing excessive information, while low-degree nodes remain under-represented.

With the social attention technique, we effectively incorporate information from socially connected items, with weights controlled by both users’ behavioral similarity and their structural properties in the social network.
This approach provides more reliable signals for user modeling in addition to the community signals of CEG.

\subsection{User-Adaptive Social Information Fusion}
\label{ssec:aggregation}
We integrate the distinct social information captured by our two modules, \modulecommunity and \moduleindividual, in a user-adaptive manner.
For each user $u$, we combine the two embeddings $\mathbf{h}_u^{\text{C}}$ and $\mathbf{h}_u^{\text{S}}$ with learnable weights, which are learned jointly with the recommendation objective.
Inspired by the gating network \cite{jacobs1991adaptive, ma2019hierarchical}, which takes disentangled embeddings as input and generates weights through a lightweight MLP, we define the adaptive weight $\alpha_u$ as follows:
\begin{equation}
    \alpha_u = \sigma_1(\sigma_2([\mathbf{h}_u^{\text{C}};\mathbf{h}_u^{\text{S}}]\mathbf{W}_1)\mathbf{W}_2),
\end{equation}
where $\mathbf{W}_1$ and $\mathbf{W}_2$ are the learnable weights of the two-layer MLP, while $\sigma_1$ and $\sigma_2$ denote activation functions set to the sigmoid and LeakyReLU functions, respectively.
Using $\alpha_u$, we integrate the two social signals adaptively for each user, as described in Equation~\eqref{eq:adaptiveagg}. This step ensures that the social information most relevant to the recommendation task---varying across users and datasets---exerts stronger influence in the final representation.

\subsection{Self-Supervised Learning for Regularizing Community Signals}
\label{ssec:ssl}

Our \modulecommunity module creates a user embedding $\mathbf{H}_u^{\text{C}}$ as a weighted average of community embeddings $\mathbf{H}_{\mathcal{C}}$ containing the user $u$, based on the user-community membership $\mathbf{G}$ identified by our overlapping community detection algorithm. As depicted by the blue and green paths on the left side of Figure~\ref{fig-ssl-detail}, the design of \method allows us to effectively combine those community-level signals and user–item interaction information in the final user representations,
but introduces two notable limitations that we aim to address through self-supervised learning.

First, for each user $u$, the information from communities that are irrelevant to $u$ is largely included in their final representation $\mathbf{e}_u$ as a result of subsequent message passing on the user-item interaction graph.
This can make user representations overly smooth compared to previous approaches that have explicit learnable embeddings for users.
Second, the community memberships may contain structural or statistical biases as they are extracted from our detection algorithm without ground-truth labels.
If such biased assignments are not aligned with the users' actual preferences, the user embeddings may incorporate conflicting community signals.



To address these limitations, we adopt contrastive learning that regularizes community-aware user representations by promoting consistency across perturbed views. The right side of Figure~\ref{fig-ssl-detail} illustrates the self-supervised learning process.
We first create two augmented views for each user by randomly masking a proportion of the non-zero entries in the user-community matrix $\mathbf{G}$. Then, we use InfoNCE \cite{oord2018representation} as the objective function, which is defined as:
\begin{equation}
     \mathcal{L}_{\text{ssl}}(\theta) = -{\sum_{u \in \mathcal{U}} \log \frac{\exp(\text{sim}(\mathbf{e}_{u}^{'}, \mathbf{e}_{u}^{''}) / \tau)}{\sum_{v\in \mathcal{U}} \exp(\text{sim}(\mathbf{e}_{u}^{'}, \mathbf{e}_{v}^{''}) / \tau)}},
\end{equation}
where $\text{sim}()$ is the cosine similarity, and $\tau$ is a temperature hyperparameter.
Here, $\mathbf{e}_{u}^{'}$ and $\mathbf{e}_{u}^{''}$ represent two independently perturbed views of user $u$’s representation.
We control the ratio of masking for creating each view as a hyperparameter $0 < \rho < 1$.

By incorporating this objective, we effectively address both limitations.
First, we control the influence of communities that are less relevant to each target user.
The InfoNCE loss maximizes the mutual information between two augmented views for each user \cite{10.1007/978-3-030-58621-8_45}. By jointly optimizing this term alongside the main recommendation loss, the model is guided to produce consistent representations centered around the user’s most relevant communities.
Second, we mitigate potential bias in $\mathbf{G}$ by exposing the model to diverse, perturbed versions of it. This encourages the model to produce consistent outputs across varying views, thereby improving its robustness to noise and instability in community memberships.

\section{Analysis and Discussion}
\label{sec:parameter-efficient}
In this section, we analyze the effectiveness of \method in terms of storage efficiency (i.e., the number of parameters). For time efficiency, we provide a detailed time-complexity analysis in Appendix~\ref{appendix:timecomplexity}. Furthermore, we present toy experimental results demonstrating the robustness of our user modeling for cold-start users.

\paragraph{Parameter-Efficient User Modeling}
\method drastically reduces the number of learnable parameters required for user modeling and eliminates the linear parameter growth with respect to the number of users.
In our design, user modeling requires only the embeddings of communities, which are far fewer in number than actual users, along with the weight parameters of the gating network.

To validate the parameter efficiency of \method, in Figure~\ref{fig-parameter-analysis}, we compare the number of parameters against LightGCN \cite{10.1145/3397271.3401063}, the most lightweight GCF model.
\method substantially reduces the number of parameters required for modeling user representations, decreasing the overall parameter size by millions.
Specifically, it achieves reductions of approximately 36\% on Douban-Book, 45\% on Yelp, and 28\% on Epinions in terms of total parameters.
This significantly lowers computational costs and improves the scalability of recommender systems especially when the number of users is large.

\begin{table}
\huge
\begin{center}
\caption{Performance comparison between \method and GCF and graph-based \SR models on cold-start users. \method generates effective user representations for these users and maintains strong recommendation performance.}
\vspace{-10pt}
\label{table-coldstart}
\resizebox{\columnwidth}{!}{
\begin{tabular}{l|cccc|cccc}
\toprule
\multirow{2}{*}{Method} & \multicolumn{4}{c|}{Douban-Book} & \multicolumn{4}{c}{Yelp} \\
\cmidrule{2-9}
& R@10~ & R@20~ & N@10~ & N@20~ & R@10~ & R@20~ & N@10~ & N@20~ \\

\midrule
LightGCN &\underline{.0247}~ &\underline{.0307}~& \underline{.0275}~ & \underline{.0274}~ & .0277~ & .0393~ & .0184~& .0219 \\
GBSR & .0161~ & .0279~ & .0236~ & .0265~ & \underline{.0613}~ & \textbf{.0933}~ & \underline{.0382}~ & \underline{.0477} \\
\midrule
\textbf{\method (ours)} & \textbf{.0387}~ & \textbf{.0617}~ & \textbf{.0499}~ & \textbf{.0542}~ & \textbf{.0666}~ & \underline{.0922}~ & \textbf{.0426}~ & \textbf{.0507}~ \\
\bottomrule

\end{tabular}
}
\end{center}
\vspace{-5pt}
\end{table}

\paragraph{Generalizable User Modeling for Cold-Start Users}
Providing accurate recommendations for cold-start users, who have not interacted with any items in the training data, is a critical yet highly challenging problem in GCF~\cite{panteli2023addressing, liang2020joint, liu2020heterogeneous}.
By incorporating recommendation-specific community signals and socially connected item signals from their friends, we generalize our model to these users.

We design a dedicated experiment to validate this point.
From the original dataset, we randomly sample 500 users as the cold-start set and remove their interaction histories from the training data.
During inference, we evaluate the recommendation performance on the 500 held-out cold-start users. We compare our method against LightGCN \cite{10.1145/3397271.3401063} and GBSR \cite{10.1145/3637528.3671807} under the same conditions on two datasets (Douban-Book and Yelp). GBSR is chosen as a representative graph-based \SR method due to its strong performance in the original setting (as shown in Section~\ref{ssec:overall-performance}).

As shown in Table~\ref{table-coldstart}, \method achieves strong performance for cold-start users compared to the baselines.
LightGCN, which relies solely on user-item interaction histories, fails to provide effective recommendations for cold-start users across both datasets. GBSR performs well on Yelp, which contains abundant social information, but falls below LightGCN on Douban-Book, where social information is relatively sparse. This suggests that its effectiveness is highly dependent on the characteristics of social information in each dataset.
By contrast, our approach leverages community information and socially-connected item embeddings, aggregated through an adaptive gating network, to generate strong and generalizable representations for cold-start users.

\begin{table}[t]
\large
\begin{center}
\caption{Dataset statistics.}
\vspace{-10pt}
\label{table-data-statistic}
\resizebox{\columnwidth}{!}{
\begin{tabular}{l|rrr}
\toprule
Dataset & Douban-Book & Yelp & Epinions \\ 
\midrule
\# of Users & 13,024 & 19,539 & 18,202 \\ 
\# of Items & 22,347 & 22,228 & 47,449 \\ 
\# of Interactions & 598,420 & 450,884 & 338,400 \\ 
Interaction Density (\%) & 0.206 & 0.104 & 0.039 \\ 
\midrule
\# of Social Relations & 169,150 & 727,384 & 595,259 \\ 
Social Relation Density (\%) & 0.100 & 0.191 & 0.180 \\
\bottomrule 
\end{tabular}
}
\end{center}
\vspace{-5pt}
\end{table}


\begin{table*}[thbp]
\begin{center}
\caption{
    Performance comparison between \method and baselines. 
    \method consistently outperforms all baselines across different datasets and evaluation metrics, demonstrating its effectiveness.
    Bold values indicate the best performance, while underlined values represent the second-best (R = Recall, N = NDCG).
    We also report the improvement of \method over the best competitor.
}
\vspace{-10pt}
\label{table-overall-performance}
\huge
\resizebox{\textwidth}{!}{
\begin{tabular}{l| cccccc | cccccc | cccccc}
\toprule
\multirow{2}{*}{Method}  & \multicolumn{6}{c|}{Douban-Book} & \multicolumn{6}{c|}{Yelp} & \multicolumn{6}{c}{Epinions} \\
\cmidrule{2-19}

 & R@10 & R@20 & R@40 & N@10 & N@20 & N@40
& R@10 & R@20 & R@40 & N@10 & N@20 & N@40
& R@10 & R@20 & R@40 & N@10 & N@20 & N@40 \\
\midrule

LightGCN & .0866 & .1311 & .1915 & .0959 & .1058 & .1244 & .0618 & .1012 & .1566 & .0448 & .0568 & .0718 & .0403 & .0619 & .0938 & .0288 & .0352 & .0437 \\
SGL & .0894 & .1328 & .1882 & .1010 & .1104 & .1277  & .0691 & .1093 & .1660 & .0500 & .0622 & .0777 & .0412 & .0640 & .0933 & .0293 & .0362 & .0441 \\ 
SimGCL & .0890 & .1321 & .1896 & .1015 & .1107 & .1286 & .0688 & .1092 & .1664 & .0501 & .0625 & .0780 & .0414 & .0653 & .0962 & .0294 & .0366 & .0449\\
LightGCL & .0800 & .1198 & .1654 & .0937 & .1020 & .1159 & .0667 & .1042 & .1581 & .0490 & .0605 & .0753& .0420 & .0634 & .0945 & .0305 & .0369 & .0452\\
Gformer &.0930&.1349&.1905&.1064&.1149&.1321&.0651 & .1046 & .1597 & .0471 & .0592 & .0742&.0386 & .0604 & .0890 & .0277 & .0343 & .0420\\
AutoCF & \underline{.0963} & .1389 & .1927 & \underline{.1142} & \underline{.1220} & \underline{.1382}& {.0707} & {.1097} & .1668 & {.0523} & {.0643} & {.0800} & .0386 & .0601 & .0895 & .0272 & .0337 & .0416\\
MixSGCL & .0956 & .1399 & .1941 & .1063 & .1155 & .1319 & \underline{.0711} & \underline{.1124} & .1680 & \underline{.0527} & \underline{.0653} & \underline{.0803} & .0419 & .0640 & .0913 & .0303 & .0369 & .0443\\
\midrule
MHCN & .0948& \underline{.1412} & \underline{.2001} & .1069 & .1165 & .1345 & .0678 & .1052 & .1605 & .0499 & .0615 & .0766 & .0420 & .0649 & .0978 & .0304 & .0373 & .0461\\
DSL$^{*}$ & .0808 & .1227 & .1776 & .0897 & .0987 & .1157 & .0576 & .0950 & .1450 & .0418 & .0534 & .0670 & .0392 & .0606 & .0900 & .0279 & .0343 & .0421 \\
GDMSR$^{*}$ & .0580 & .0903 & .1406 & .0652 & .0724 & .0875 & .0501 & .0824 & .1312 & .0353 & .0453 & .0584 & .0308 & .0499 & .0768 & .0218 & .0275 & .0345 \\
GBSR & .0935 & .1352 & .1840 & .1092 & .1167 & .1306 & .0690 & .1094 & \underline{.1684} & .0494 & .0618 & .0777 & \underline{.0446} & \underline{.0671} & \underline{.1001} & \underline{.0320} & \underline{.0388} & \underline{.0476} \\
RecDiff$^{*}$ & .0736 & .1165 & .1724 & .0830 & .0933 & .1107 & .0557 & .0924 & .1457 & .0398 & .0510 & .0655 & .0315 & .0507 & .0800 & .0220 & .0279 & .0358 \\

SGIL & .0917 & .1350 & .1885  & .1067 & .1146 & .1303 & .0628 & .0977 & .1491 & .0468 & .0576 & .0717 & .0408 & .0633 & .0939 & .0288 & .0355 & .0438  \\

\midrule
\multirow{2}{*}{\textbf{\method (ours)}} & \textbf{.1079} & \textbf{.1562} & \textbf{.2146} & \textbf{.1238} &\textbf{.1330} & \textbf{.1502} & \textbf{.0773} & \textbf{.1214} & \textbf{.1818} & \textbf{.0572} & \textbf{.0706} & \textbf{.0869}
& \textbf{.0478} & \textbf{.0727} & \textbf{.1061} & \textbf{.0347} & \textbf{.0421} & \textbf{.0510} \\
 & +12.0\% & +10.6\% & +7.2\% & +8.4\% & +9.0\% & +8.7\%
& +8.7\% & +8.0\% & +8.0\% & +8.5\% & +8.1\% & +8.2\%
& +7.2\% & +8.3\% & +6.0\% & +8.4\% & +8.5\% & +7.1\% \\

\bottomrule 
\end{tabular}
}
\end{center}

\footnotesize
\begin{tabular}{@{}p{\linewidth}@{}}
${*}$ The experimental settings differ from those in the original papers, including the use of validation data for better reproducibility. 
See Appendix \ref{appendix-exp-details} for details.
\end{tabular}

\vspace{-5pt}
\end{table*}



\section{Experiments}
\label{sec:experiments}

  

  
  

We conduct comprehensive experiments to empirically validate the effectiveness of \method.
In addition to the results presented in this section (\S\ref{ssec:overall-performance} - \S\ref{ssec:robustness-to-interaction-degrees}), we provide ablation studies, hyperparameter sensitivity analyses, comparisons with parameter-efficient LightGCN variants, and 
robustness to adversarial noise in Appendix~\ref{appendix:ablation}.

\subsection{Experimental Settings}

We run all experiments using SSLRec \cite{ren2024sslrec}, a popular framework that provides implementations of diverse baselines and datasets.

\paragraph{Datasets}
We use three popular datasets for social recommendation: Douban-Book\footnote{\url{https://github.com/librahu/HIN-Datasets-for-Recommendation-and-Network-Embedding}}, Yelp\footnote{\url{https://github.com/Coder-Yu/QRec}}, and Epinions\footnote{\url{http://www.trustlet.org/downloaded epinions.html}}, following \cite{10.1145/3637528.3671807}.
Each dataset is divided into training, validation, and test sets in a 6:2:2 ratio. 
Their statistics are given in Table \ref{table-data-statistic}.

\paragraph{Baselines}
We compare \method against six state-of-the-art graph-based \SR models and seven advanced GCF approaches.

For GCF baselines, 
LightGCN \cite{10.1145/3397271.3401063} simplifies graph convolution by removing feature transformation and nonlinear activation.
SGL \cite{wu2021self} introduces contrastive learning with randomly perturbed graph views, while SimGCL \cite{yu2022graph} performs lightweight contrastive regularization through stochastic embedding perturbations. 
LightGCL \cite{cai2023lightgclsimpleeffectivegraph} constructs augmentation-free spectral views via singular value decomposition for global structural refinement. 
AutoCF \cite{10.1145/3543507.3583336} and GFormer \cite{10.1145/3539618.3591723} extend the masked autoencoder paradigm to graph recommendation: the former introduces structure-adaptive augmentation, whereas the latter captures invariant collaborative rationales inspired by graph transformers.
MixSGCL \cite{zhang2024mixedsupervisedgraphcontrastive} further performs node- and edge-level mixup to synthesize informative contrastive views and alleviate gradient inconsistency.

For \SR models, 
MHCN \cite{10.1145/3442381.3449844} captures higher-order social ties via hypergraph convolution coupled with self-supervision, and DSL \cite{wang2023denoisedselfaugmentedlearningsocial} aligns interaction- and socially driven semantics through contrastive learning. 
GDMSR \cite{quan2023robust} introduces a preference-guided denoising mechanism with a self-correcting curriculum, while GBSR \cite{10.1145/3637528.3671807} employs the information bottleneck principle to obtain a minimal yet sufficient denoised social graph. 
RecDiff \cite{10.1145/3627673.3679630} reconstructs reliable social signals via a hidden-space diffusion process, and SGIL 
\cite{10.1145/3726302.3730013}
further refines this idea by introducing latent diffusion for iterative noise propagation and relation recovery.
\paragraph{Hyperparameters}
All learnable parameters are initialized using Xavier initialization \cite{glorot2010understanding}.
We train all models using the Adam optimizer \cite{kingma2014adam} with a fixed learning rate of $10^{-3}$ and a batch size of 4096.
For fair and rigorous evaluation, hyperparameter configurations for all models, including \method, are determined through extensive grid search, with reference to the original papers of each method. 
\paragraph{Evaluation Metrics}
We evaluate model performance using two popular ranking-based metrics: Recall@$K$ and NDCG@$K$ \cite{gunawardana2009survey, steck2013evaluation}, where $K$ is set to 10, 20, and 40.
Each model is trained for up to 500 epochs with early stopping (patience = 15) based on NDCG@20, and results are reported from a single evaluation run.
For evaluation, we consider all non-interacted items in the training set for each user as test candidates for a fair evaluation \cite{zhao2020revisiting}.

\subsection{Overall Performance}
\label{ssec:overall-performance}

Table~\ref{table-overall-performance} shows the performance of \method against state-of-the-art baseline models on the three benchmark datasets. 
\method consistently outperforms all baselines, including advanced GCF and graph-based \SR methods. Specifically, \method achieves improvements of 9.0\%, 8.1\%, and 8.5\% in NDCG@20 over the second-best competitors across all datasets.

Notably, \method effectively mitigates the sparsity of user-item interactions by leveraging the relatively denser structure of the social network. This effect is particularly evident on the Epinions dataset, where user-item interactions are highly sparse while the social network is comparatively dense. In such settings, \method is able to capture robust social community signals and diverse socially connected item signals, leading to more informative user representations and, consequently, improved recommendation accuracy.




\begin{figure}
\centering
\includegraphics[width=0.9\linewidth]{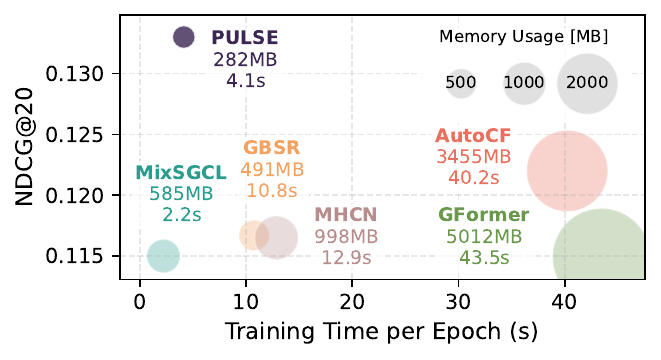}

\vspace{-10pt}
\caption{
Training efficiency analysis on the Douban-Book dataset. \method achieves both minimal memory usage and fast training, outperforming baselines in overall efficiency.
}
\Description[Training efficiency comparison of recommendation models]{
Bubble scatter plot showing the relationship between training time, performance, and memory usage for several recommendation models on the Douban-Book dataset. 
The horizontal axis represents training time per epoch in seconds, and the vertical axis represents normalized discounted cumulative gain at rank twenty. 
Each model is shown as a labeled circle, where larger circles indicate higher memory usage measured in megabytes. 
PULSE is located near the upper-left area of the plot, combining the shortest training time with the highest performance and the smallest memory footprint. 
MixSGCL and GBSR appear at low training times with slightly lower performance and larger memory usage than PULSE. 
MHCN is positioned at moderate training time and performance with increased memory usage. 
AutoCF and GFormer are located at the highest training times and have the largest circles, indicating substantially higher memory consumption.
}
\vspace{-5pt}
\label{fig-efficiency-analysis}
\end{figure}
\subsection{Training Efficiency of \method}
\label{ssec:training-efficiency}
To evaluate the computational efficiency of \method, we measure per-epoch training time and maximum GPU memory usage during training on the Douban-Book dataset. As shown in Figure~\ref{fig-efficiency-analysis}, \method achieves minimal resource consumption while maintaining fast training. Specifically, it requires only 4.1 seconds per epoch with a memory footprint of 282 MB, making it the most memory-efficient method among all baselines. 
In contrast, advanced GCF methods such as AutoCF (40.2s, 3455 MB) and GFormer (43.5s, 5012 MB) incur substantial computational overhead, while MixSGCL (2.2s, 585 MB), despite faster training, consumes significantly more memory. Graph-based \SR baselines, including GBSR (10.8s, 491 MB) and MHCN (12.9s, 998 MB), also lag behind \method in both training efficiency and memory usage.
Furthermore, in terms of convergence speed, \method reaches its best performance in an average of 349 seconds across three datasets, which is comparable to MixSGCL (213s) and MHCN (316s), and substantially faster than GBSR (608s), AutoCF (2507s), and GFormer (1461s).
These results demonstrate that \method enables cost-effective user representation learning, highlighting its practicality for large-scale recommendation.



\begin{figure}[t]
\centering
\includegraphics[width=\columnwidth]{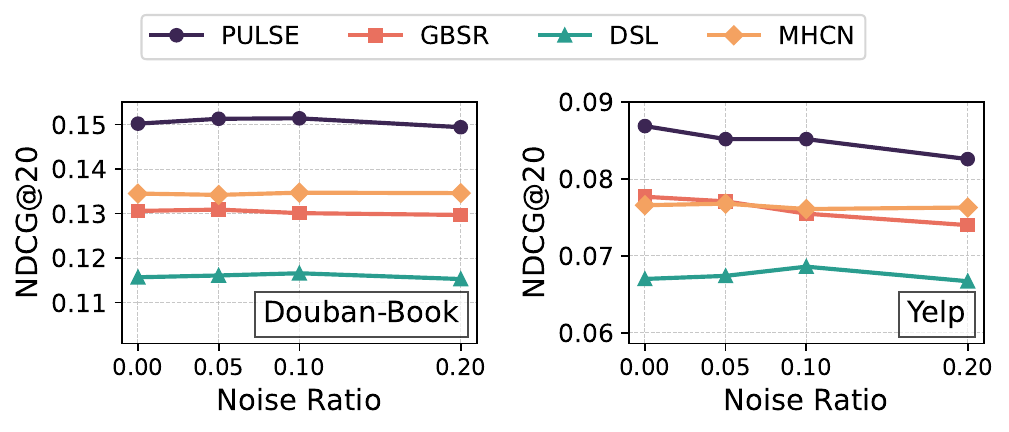}
\vspace{-20pt}
\caption{
    Performance under varying levels of social noise on Douban-Book and Yelp. \method consistently demonstrates strong robustness to increasing noise.
}
\label{fig-noise-robustness}
\Description[Effect of social noise on recommendation performance]{
Two line charts showing recommendation performance under increasing levels of social noise for the Douban-Book dataset on the left and the Yelp dataset on the right.
In both charts, the horizontal axis represents noise ratio from zero to zero point two, and the vertical axis represents normalized discounted cumulative gain at rank twenty.
Four methods are plotted as separate lines: PULSE, GBSR, DSL, and MHCN.
On Douban-Book, PULSE maintains the highest performance across all noise levels with only a slight decrease as noise increases, while the other methods show lower performance with minor fluctuations.
On Yelp, PULSE again shows the highest values and a gradual decline with increasing noise, whereas GBSR and MHCN show small decreases and DSL remains consistently lower.
Overall, the plots show that PULSE is less sensitive to increasing social noise compared to the other methods on both datasets.
}
\vspace{-5pt}
\end{figure}
\subsection{Robustness to Social Noise}
\label{ssec:robustness-to-social-noise}

\method exhibits strong resilience to noise in the social network. 
To evaluate this, we design an experiment in which noise is artificially injected into the social network. 
Specifically, for the Douban-Book and Yelp datasets, we randomly remove
5\%, 10\%, and 20\% of the original social edges and replace them with randomly
generated ones.
We then compare the performance 
against the three strong baselines, GBSR, DSL, and MHCN.
As shown in Figure~\ref{fig-noise-robustness}, \method consistently maintains the strongest performance as the noise level increases.
We attribute this resilience to its design: community-level signals, which are less sensitive to noisy connections, and socially connected items, whose influence is modulated by behavioral similarity between users, thereby ignoring spurious links. 

\begin{figure}
\centering
\includegraphics[width=\columnwidth]{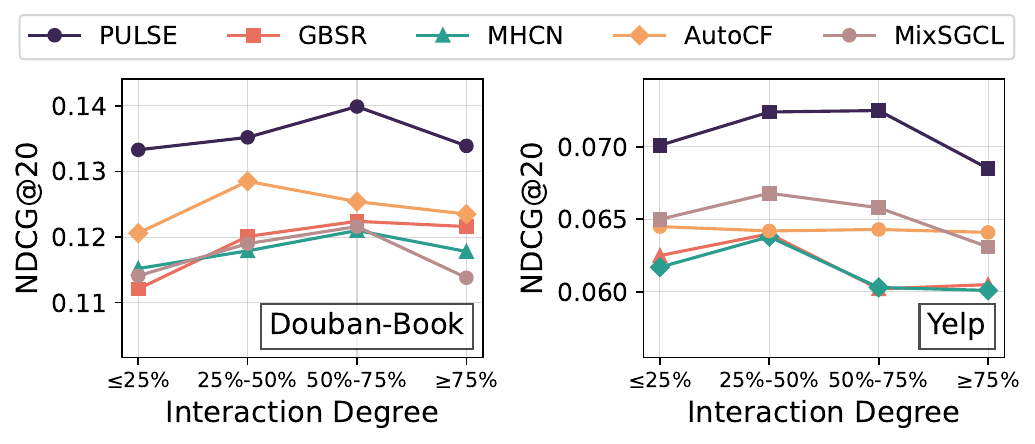}
\vspace{-20pt}
\caption{Performance across different interaction degree groups on Douban-Book and Yelp. \method consistently outperforms baselines across all groups in both datasets.}

\Description[Performance across interaction degree groups]{
Two line charts showing recommendation performance across different interaction degree groups for the Douban-Book dataset on the left and the Yelp dataset on the right.
In both charts, the horizontal axis represents interaction degree groups ranging from less than or equal to twenty five percent to greater than or equal to seventy five percent, and the vertical axis represents normalized discounted cumulative gain at rank twenty.
Five methods are plotted as separate lines: PULSE, GBSR, MHCN, AutoCF, and MixSGCL.
On Douban-Book, PULSE achieves the highest performance across all interaction degree groups, with a peak in the middle degree range, while the baseline methods show lower values with smaller variations.
On Yelp, PULSE again shows the highest performance across all degree groups, while the other methods exhibit flatter trends and lower overall performance.
The plots indicate that PULSE maintains strong performance consistency across both sparse and dense interaction groups compared to the baseline methods.
}
\label{fig-robustness-sparsity}
\vspace{-5pt}
\end{figure}
\subsection{Robustness to Interaction Degrees}
\label{ssec:robustness-to-interaction-degrees}

The superior performance of \method is observed not only for users with certain interaction degrees, but consistently across all users. To assess the robustness of the social signals leveraged by \method under varying interaction degrees, we conduct experiments by dividing users into four groups based on their interaction degree percentiles: 0–25\%, 25–50\%, 50–75\%, and 75–100\%. Performance is evaluated on Douban-Book and Yelp, as shown in Figure~\ref{fig-robustness-sparsity}.
\method consistently outperforms the strongest baselines across all interaction degree levels. Notably, even in the lowest interaction-degree group where user interactions are highly limited (Douban-Book: fewer than 8 interactions; Yelp: fewer than 4 interactions), \method maintains stable and strong performance on both datasets. This shows that the user representations generated by our approach capture meaningful social signals for recommendation more effectively than other methods that incorporate social information do.
\section{Conclusion}
\label{sec:conclusion}

In this paper, we introduced \method, a novel approach enhancing the efficiency of graph collaborative filtering by reducing the reliance on explicit user embeddings through the use of socially derived information. \method leverages two types of meaningful social signals, communities to which users belong and items interacted with by their social neighbors, and integrates them effectively and adaptively via a gating network for user representation modeling. This design achieves superior performance, 
while requiring only a small fraction of the parameters used for user modeling in standard GCF settings.
For future work, we plan to further improve the efficiency of GCF by extending our approach to reduce the cost of item embeddings.
Beyond efficiency, we further extend the self-supervised objective to enable more robust representation learning under diverse noisy conditions in both interaction and social relations, beyond community-level structures.
Additionally, we aim to broaden our parameter-efficient modeling framework by incorporating diverse meta features beyond social recommendation.
\begin{acks}
    This work was supported by the National Research Foundation of Korea (NRF) grant funded by the Korea government (MSIT) (RS-2024-00341425 and RS-2024-00406985).
    Jaemin Yoo is the corresponding author.
\end{acks}

\clearpage
\bibliographystyle{ACM-Reference-Format}
\bibliography{references}

@inproceedings{panteli2023addressing,
  author    = {Panteli, Antiopi and Boutsinas, Basilis},
  title     = {Addressing the cold-start problem in recommender systems based on frequent patterns},
  booktitle = {Algorithms},
  pages     = {182},
  year      = {2023}
}

@article{choi2025simple,
  title={Simple and Behavior-Driven Augmentation for Recommendation with Rich Collaborative Signals},
  author={Choi, Doyun and Lee, Cheonwoo and Yoo, Jaemin},
  journal={arXiv preprint arXiv:2511.00436},
  year={2025}
}

@inproceedings{10.1145/1458082.1458205,
  author    = {Ma, Hao and Yang, Haixuan and Lyu, Michael R. and King, Irwin},
  title     = {Sorec: social recommendation using probabilistic matrix factorization},
  booktitle = {CIKM},
  year      = {2008}
}

@inproceedings{10.1145/3627673.3679630,
  author    = {Li, Zongwei and Xia, Lianghao and Huang, Chao},
  title     = {Recdiff: Diffusion Model for Social Recommendation},
  booktitle = {CIKM},
  year      = {2024}
}

@inproceedings{10.1145/3397271.3401063,
  author    = {He, Xiangnan and Deng, Kuan and Wang, Xiang and Li, Yan and Zhang, YongDong and Wang, Meng},
  title     = {Lightgcn: Simplifying and Powering Graph Convolution Network for Recommendation},
  booktitle = {SIGIR},
  year      = {2020}
}

@inproceedings{sahebi2011community,
  author    = {Sahebi, Shaghayegh and Cohen, William W},
  title     = {Community-based recommendations: a solution to the cold start problem},
  booktitle = {RSWEB},
  year      = {2011}
}

@misc{rendle2012bprbayesianpersonalizedranking,
      title={BPR: Bayesian Personalized Ranking from Implicit Feedback}, 
      author={Steffen Rendle and Christoph Freudenthaler and Zeno Gantner and Lars Schmidt-Thieme},
howpublished = {arXiv preprint arXiv:1205.2618},
      year={2012},
}

@inproceedings{ren2024sslrec,
  author    = {Ren, Xubin and Xia, Lianghao and Yang, Yuhao and Wei, Wei and Wang, Tianle and Cai, Xuheng and Huang, Chao},
  title     = {Sslrec: A self-supervised learning framework for recommendation},
  booktitle = {WSDM},
  year      = {2024}
}

@article{chen2024dual,
  title={Dual-Domain Collaborative Denoising for Social Recommendation},
  author={Chen, Wenjie and Zhang, Yi and Li, Honghao and Sang, Lei and Zhang, Yiwen},
  journal={IEEE Transactions on Computational Social Systems},
  year={2025},
  publisher={IEEE}
}

@article{annurev:/content/journals/10.1146/annurev.soc.27.1.415,
   author = "McPherson, Miller and Smith-Lovin, Lynn and Cook, James M",
   title = "Birds of a Feather: Homophily in Social Networks", 
   journal= "Annual Review of Sociology",
   year = "2001",
   volume = "27",
   number = "Volume 27, 2001",
   pages = "415-444",
   publisher = "Annual Reviews",
   issn = "1545-2115",
   type = "Journal Article",
  }

@inproceedings{hogg2004enhancing,
  author    = {Hogg, Tad and Adamic, Lada},
  title     = {Enhancing reputation mechanisms via online social networks},
  booktitle = {EC},
  year      = {2004}
}

@article{doi:10.1177/0049124193022001006,
  author    = {PETER V. MARSDEN and NOAH E. FRIEDKIN},
  title     = {Network Studies of Social Influence},
  journal = {Sociological Methods \& Research},
  pages     = {127--151},
  year      = {1993}
}

@article{annurev:/content/journals/10.1146/annurev.psych.55.090902.142015,
   author = "Cialdini, Robert B. and Goldstein, Noah J.",
   title = "Social Influence: Compliance and Conformity", 
   journal= "Annual Review of Psychology",
   year = "2004",
   volume = "55",
   number = "Volume 55, 2004",
   pages = "591-621",
   publisher = "Annual Reviews",
   issn = "1545-2085",
   type = "Journal Article",
  }

@inproceedings{10.1145/1864708.1864736,
  author    = {Jamali, Mohsen and Ester, Martin},
  title     = {A matrix factorization technique with trust propagation for recommendation in social networks},
  booktitle = {RecSys},
  year      = {2010}
}

@inproceedings{Guo_Zhang_Yorke-Smith_2015,
  author    = {Guo, Guibing and Zhang, Jie and Yorke-Smith, Neil},
  title     = {Trustsvd: Collaborative Filtering with Both the Explicit and Implicit Influence of User Trust and of Item Ratings},
  booktitle = {AAAI},
  pages     = {},
  year      = {2015}
}

@inproceedings{10.1145/3331184.3331214,
  author    = {Wu, Le and Sun, Peijie and Fu, Yanjie and Hong, Richang and Wang, Xiting and Wang, Meng},
  title     = {A Neural Influence Diffusion Model for Social Recommendation},
  booktitle = {SIGIR},
  year      = {2019}
}

@inproceedings{10.1145/3357384.3357924,
  author    = {Xu, Fengli and Lian, Jianxun and Han, Zhenyu and Li, Yong and Xu, Yujian and Xie, Xing},
  title     = {Relation-aware Graph Convolutional Networks for Agent-Initiated Social E-Commerce Recommendation},
  booktitle = {CIKM},
  year      = {2019}
}

@article{LIAO2022595,
  author    = {Jie Liao and Wei Zhou and Fengji Luo and Junhao Wen and Min Gao and Xiuhua Li and Jun Zeng},
  title     = {Sociallgn: Light graph convolution network for social recommendation},
  journal = {Information Sciences},
  pages     = {595--607},
  year      = {2022}
}

@inproceedings{wang2023denoisedselfaugmentedlearningsocial,
      title={Denoised Self-Augmented Learning for Social Recommendation}, 
      author={Tianle Wang and Lianghao Xia and Chao Huang},
    booktitle={IJCAI},
      year={2023}
}

@inproceedings{10.1145/3442381.3449844,
  author    = {Yu, Junliang and Yin, Hongzhi and Li, Jundong and Wang, Qinyong and Hung, Nguyen Quoc Viet and Zhang, Xiangliang},
  title     = {Self-supervised Multi-Channel Hypergraph Convolutional Network for Social Recommendation},
  booktitle = {WWW},
  year      = {2021}
}

@inproceedings{10.1145/3447548.3467340,
  author    = {Yu, Junliang and Yin, Hongzhi and Gao, Min and Xia, Xin and Zhang, Xiangliang and Viet Hung, Nguyen Quoc},
  title     = {Socially-aware Self-Supervised Tri-Training for Recommendation},
  booktitle = {KDD},
  year      = {2021}
}

@inproceedings{10.1145/3589334.3645460,
  author    = {Jiang, Wei and Gao, Xinyi and Xu, Guandong and Chen, Tong and Yin, Hongzhi},
  title     = {Challenging Low Homophily in Social Recommendation},
  booktitle = {WWW},
  year      = {2024}
}

@inproceedings{10.1145/3637528.3671807,
  author    = {Yang, Yonghui and Wu, Le and Wang, Zihan and He, Zhuangzhuang and Hong, Richang and Wang, Meng},
  title     = {Graph Bottlenecked Social Recommendation},
  booktitle = {KDD},
  year      = {2024}
}

@inproceedings{10.1145/3637528.3671958,
  author    = {Sun, Youchen and Sun, Zhu and Du, Yingpeng and Zhang, Jie and Ong, Yew Soon},
  title     = {Self-supervised Denoising through Independent Cascade Graph Augmentation for Robust Social Recommendation},
  booktitle = {KDD},
  year      = {2024}
}

@inproceedings{wu2021self,
  author    = {Wu, Jiancan and Wang, Xiang and Feng, Fuli and He, Xiangnan and Chen, Liang and Lian, Jianxun and Xie, Xing},
  title     = {Self-supervised graph learning for recommendation},
  booktitle = {SIGIR},
  year      = {2021}
}

@inproceedings{yu2022graph,
  author    = {Yu, Junliang and Yin, Hongzhi and Xia, Xin and Chen, Tong and Cui, Lizhen and Nguyen, Quoc Viet Hung},
  title     = {Are graph augmentations necessary? simple graph contrastive learning for recommendation},
  booktitle = {SIGIR},
  year      = {2022}
}

@inproceedings{quan2023robust,
  author    = {Quan, Yuhan and Ding, Jingtao and Gao, Chen and Yi, Lingling and Jin, Depeng and Li, Yong},
  title     = {Robust preference-guided denoising for graph based social recommendation},
  booktitle = {WWW},
  year      = {2023}
}

@inproceedings{glorot2010understanding,
  author    = {Glorot, Xavier and Bengio, Yoshua},
  title     = {Understanding the difficulty of training deep feedforward neural networks},
  booktitle = {AISTATS},
  year      = {2010}
}

@inproceedings{kingma2014adam,
  author       = {Diederik P. Kingma and
                  Jimmy Ba},
  title        = {Adam: {A} Method for Stochastic Optimization},
  booktitle    = {ICLR},
  year         = {2015}
}

@inproceedings{zhao2020revisiting,
  author    = {Zhao, Wayne Xin and Chen, Junhua and Wang, Pengfei and Gu, Qi and Wen, Ji-Rong},
  title     = {Revisiting alternative experimental settings for evaluating top-n item recommendation algorithms},
  booktitle = {CIKM},
  year      = {2020}
}

@article{li2022disentangled,
  author    = {Li, Nian and Gao, Chen and Jin, Depeng and Liao, Qingmin},
  title     = {Disentangled modeling of social homophily and influence for social recommendation},
  journal = {IEEE Transactions on Knowledge and Data Engineering},
  pages     = {5738--5751},
  year      = {2022}
}

@article{hu2024hierarchical,
  author={Hu, Zheng and Nakagawa, Satoshi and Zhuang, Yan and Deng, Jiawen and Cai, Shimin and Zhou, Tao and Ren, Fuji},
journal={ IEEE Transactions on Knowledge and Data Engineering },
title={ Hierarchical Denoising for Robust Social Recommendation },
year={2025},
pages={739--753},
}

@inproceedings{liu2024learning,
  author    = {Liu, Nian and Fan, Shen and Bai, Ting and Wang, Peng and Sun, Mingwei and Mo, Yanhu and Xu, Xiaoxiao and Liu, Hong and Shi, Chuan},
  title     = {Learning Social Graph for Inactive User Recommendation},
  booktitle = {DASFAA},
  year      = {2024}
}

@inproceedings{hu2023celebrity,
  author    = {Hu, Zheng and Nakagawa, Satoshi and Luo, Liang and Gu, Yu and Ren, Fuji},
  title     = {Celebrity-aware Graph Contrastive Learning Framework for Social Recommendation},
  booktitle = {CIKM},
  year      = {2023}
}

@inproceedings{zhao2014leveraging,
  author    = {Zhao, Tong and McAuley, Julian and King, Irwin},
  title     = {Leveraging social connections to improve personalized ranking for collaborative filtering},
  booktitle = {CIKM},
  year      = {2014}
}

@article{gunawardana2009survey,
  author    = {Gunawardana, Asela and Shani, Guy},
  title     = {A survey of accuracy evaluation metrics of recommendation tasks.},
  journal = {Journal of Machine Learning Research},
  pages     = {2935–2962},
  year      = {2009}
}

@inproceedings{steck2013evaluation,
  author    = {Steck, Harald},
  title     = {Evaluation of recommendations: rating-prediction and ranking},
  booktitle = {RecSys},
  year      = {2013}
}

@inproceedings{schafer1999recommender,
  author    = {Schafer, J Ben and Konstan, Joseph and Riedl, John},
  title     = {Recommender systems in e-commerce},
  booktitle = {EC},
  year      = {1999}
}

@article{lancichinetti2010characterizing,
  title={Characterizing the community structure of complex networks},
  author={Lancichinetti, Andrea and Kivel{\"a}, Mikko and Saram{\"a}ki, Jari and Fortunato, Santo},
  journal={PloS one},
  volume={5},
  number={8},
  pages={e11976},
  year={2010},
  publisher={Public Library of Science San Francisco, USA}
}

@article{gasparetti2021community,
  author    = {Gasparetti, Fabio and Sansonetti, Giuseppe and Micarelli, Alessandro},
  title     = {Community detection in social recommender systems: a survey},
  journal = {Applied Intelligence},
  pages     = {3975--3995},
  year      = {2021}
}

@inproceedings{tang2016recommendation,
  author    = {Tang, Jiliang and Wang, Suhang and Hu, Xia and Yin, Dawei and Bi, Yingzhou and Chang, Yi and Liu, Huan},
  title     = {Recommendation with social dimensions},
  booktitle = {AAAI},
  pages     = {},
  year      = {2016}
}

@inproceedings{li2015overlapping,
  author    = {Li, Hui and Wu, Dingming and Tang, Wenbin and Mamoulis, Nikos},
  title     = {Overlapping community regularization for rating prediction in social recommender systems},
  booktitle = {RecSys},
  year      = {2015}
}

@article{liu2019social,
  author    = {Liu, Huafeng and Jing, Liping and Yu, Jian and Ng, Michael K},
  title     = {Social recommendation with learning personal and social latent factors},
  journal = {IEEE Transactions on Knowledge and Data Engineering},
  pages     = {2956--2970},
  year      = {2019}
}

@article{guan2021community,
  author    = {Guan, Jiewen and Huang, Xin and Chen, Bilian},
  title     = {Community-aware social recommendation: A unified SCSVD framework},
  journal = {IEEE Transactions on Knowledge and Data Engineering},
  pages     = {2379--2393},
  year      = {2021}
}

@article{ni2023community,
  author    = {Ni, Xuelian and Xiong, Fei and Pan, Shirui and Wu, Jia and Wang, Liang and Chen, Hongshu},
  title     = {Community preserving social recommendation with Cyclic Transfer Learning},
  journal = {ACM Transactions on Information Systems},
  pages     = {1--36},
  year      = {2023}
}

@inproceedings{ni2024graph,
  author    = {Ni, Xuelian and Xiong, Fei and Zheng, Yu and Wang, Liang},
  title     = {Graph Contrastive Learning with Kernel Dependence Maximization for Social Recommendation},
  booktitle = {WWW},
  year      = {2024}
}

@inproceedings{10.1145/3331184.3331267,
  author    = {Wang, Xiang and He, Xiangnan and Wang, Meng and Feng, Fuli and Chua, Tat-Seng},
  title     = {Neural Graph Collaborative Filtering},
  booktitle = {SIGIR},
  year      = {2019}
}

@inproceedings{10.1007/978-3-030-58621-8_45,
author="Tian, Yonglong
and Krishnan, Dilip
and Isola, Phillip",
title="Contrastive Multiview Coding",
booktitle="ECCV",
year="2020",
}

@inproceedings{hu2008collaborative,
  author    = {Hu, Yifan and Koren, Yehuda and Volinsky, Chris},
  title     = {Collaborative filtering for implicit feedback datasets},
  booktitle = {ICDM},
  year      = {2008}
}

@article{linden2003amazon,
  title={Amazon.com recommendations: Item-to-item collaborative filtering},
  author={Linden, Greg and Smith, Brent and York, Jeremy},
  journal={IEEE Internet computing},
  pages={76--80},
  year={2003},
}

@article{gomez2015netflix,
  title={The netflix recommender system: Algorithms, business value, and innovation},
  author={Gomez-Uribe, Carlos A and Hunt, Neil},
  journal={ACM Transactions on Management Information Systems},
  volume={6},
  number={4},
  pages={1--19},
  year={2015},
}

@article{harper2015movielens,
  title={The movielens datasets: History and context},
  author={Harper, F Maxwell and Konstan, Joseph A},
  journal={Acm transactions on interactive intelligent systems},
  pages={1--19},
  year={2015}
}

@inproceedings{abel2011analyzing,
  title={Analyzing temporal dynamics in twitter profiles for personalized recommendations in the social web},
  author={Abel, Fabian and Gao, Qi and Houben, Geert-{Jan} and Tao, Ke},
  booktitle={WebSci},
  year={2011}
}

@inproceedings{abel2013twitter,
  title={Twitter-Based User Modeling for News Recommendations.},
  author={Abel, Fabian and Gao, Qi and Houben, Geert-{Jan} and Tao, Ke},
  booktitle={IJCAI},
  year={2013},
}

@misc{oord2018representation,
  title={Representation learning with contrastive predictive coding},
  author={Aaron van den Oord and Yazhe Li and Oriol Vinyals},
  howpublished={arXiv preprint arXiv:1807.03748},
  year={2018}
}

@inproceedings{10.1145/3726302.3730013,
author = {Yang, Yonghui and Wu, Le and Liao, Yuxin and He, Zhuangzhuang and Shao, Pengyang and Hong, Richang and Wang, Meng},
title = {Invariance Matters: Empowering Social Recommendation via Graph Invariant Learning},
year = {2025},
booktitle = {SIGIR},
}

@misc{hieu2024overlappingcommunitydetectionalgorithms,
      title={Overlapping community detection algorithms using Modularity and the cosine}, 
      author={Do Duy Hieu and Phan Thi Ha Duong},
      year={2024},
      eprint={2403.08000},
      archivePrefix={arXiv}, 
}

@article{Traag_2019,
   title={From Louvain to Leiden: guaranteeing well-connected communities},
   volume={9},
   ISSN={2045-2322},
   number={1},
   journal={Scientific Reports},
   publisher={Springer Science and Business Media LLC},
   author={Traag, V. A. and Waltman, L. and van Eck, N. J.},
   year={2019},
   }

@article{kipf2016semi,
  title={Semi-supervised classification with graph convolutional networks},
  author={Kipf, TN},
  journal={arXiv preprint arXiv:1609.02907},
  year={2016}
}

@article{xu2018powerful,
  title={How powerful are graph neural networks?},
  author={Xu, Keyulu and Hu, Weihua and Leskovec, Jure and Jegelka, Stefanie},
  journal={arXiv preprint arXiv:1810.00826},
  year={2018}
}

@article{jacobs1991adaptive,
  title={Adaptive mixtures of local experts},
  author={Jacobs, Robert A and Jordan, Michael I and Nowlan, Steven J and Hinton, Geoffrey E},
  journal={Neural computation},
  volume={3},
  number={1},
  pages={79--87},
  year={1991},
  publisher={MIT Press}
}

@inproceedings{ma2019hierarchical,
  title={Hierarchical gating networks for sequential recommendation},
  author={Ma, Chen and Kang, Peng and Liu, Xue},
  booktitle={KDD},
  year={2019}
}

@inproceedings{liang2020joint,
  title={Joint training capsule network for cold start recommendation},
  author={Liang, Tingting and Xia, Congying and Yin, Yuyu and Yu, Philip S},
  booktitle={SIGIR},
  year={2020}
}

@inproceedings{liu2020heterogeneous,
  title={A heterogeneous graph neural model for cold-start recommendation},
  author={Liu, Siwei and Ounis, Iadh and Macdonald, Craig and Meng, Zaiqiao},
  booktitle={SIGIR},
  year={2020}
}

@article{chen2024multi,
  title={Multi-view graph contrastive learning for social recommendation},
  author={Chen, Rui and Chen, Jialu and Gan, Xianghua},
  journal={Scientific reports},
  volume={14},
  number={1},
  pages={22643},
  year={2024},
  publisher={Nature Publishing Group UK London}
}

@article{ma2024robust,
  title={Robust social recommendation based on contrastive learning and dual-stage graph neural network},
  author={Ma, Gang-Feng and Yang, Xu-Hua and Long, Haixia and Zhou, Yanbo and Xu, Xin-Li},
  journal={Neurocomputing},
  volume={584},
  pages={127597},
  year={2024},
  publisher={Elsevier}
}

@misc{cai2023lightgclsimpleeffectivegraph,
      title={LightGCL: Simple Yet Effective Graph Contrastive Learning for Recommendation}, 
      author={Xuheng Cai and Chao Huang and Lianghao Xia and Xubin Ren},
      year={2023},
      eprint={2302.08191},
      archivePrefix={arXiv},
}

@inproceedings{10.1145/3543507.3583336,
author = {Xia, Lianghao and Huang, Chao and Huang, Chunzhen and Lin, Kangyi and Yu, Tao and Kao, Ben},
title = {Automated Self-Supervised Learning for Recommendation},
year = {2023},
booktitle = {WWW},
}

@inproceedings{10.1145/3539618.3591723,
author = {Li, Chaoliu and Xia, Lianghao and Ren, Xubin and Ye, Yaowen and Xu, Yong and Huang, Chao},
title = {Graph Transformer for Recommendation},
year = {2023},
booktitle = {SIGIR},
}

@misc{zhang2024mixedsupervisedgraphcontrastive,
      title={Mixed Supervised Graph Contrastive Learning for Recommendation}, 
      author={Weizhi Zhang and Liangwei Yang and Zihe Song and Henry Peng Zou and Ke Xu and Yuanjie Zhu and Philip S. Yu},
      howpublished = {arXiv preprint arXiv:2404.15954},
  year         = {2024}
}

@inproceedings{NEURIPS2024_c9e20f70,
 author = {Iyer, Roshni G. and Wang, Yewen and Wang, Wei and Sun, Yizhou},
 booktitle = {NeurIPS},
 title = {Non-Euclidean Mixture Model for Social Network Embedding},
 year = {2024}
}

@misc{zhang2023sharklightweightmodelcompression,
      title={SHARK: A Lightweight Model Compression Approach for Large-scale Recommender Systems}, 
      author={Beichuan Zhang and Chenggen Sun and Jianchao Tan and Xinjun Cai and Jun Zhao and Mengqi Miao and Kang Yin and Chengru Song and Na Mou and Yang Song},
      year={2023},
      eprint={2308.09395},
      archivePrefix={arXiv}, 
}

@inproceedings{liu2019rem,
  title={REM: From structural entropy to community structure deception},
  author={Liu, Yiwei and Liu, Jiamou and Zhang, Zijian and Zhu, Liehuang and Li, Angsheng},
  booktitle={NeurIPS},
  year={2019}
}

@inproceedings{hu2022lora,
  title={Lora: Low-rank adaptation of large language models.},
  author={Hu, Edward J and Shen, Yelong and Wallis, Phillip and Allen-Zhu, Zeyuan and Li, Yuanzhi and Wang, Shean and Wang, Lu and Chen, Weizhu and others},
  booktitle={ICLR},
  year={2022}
}

@inproceedings{
liu2021learnable,
title={Learnable Embedding sizes for Recommender Systems},
author={Siyi Liu and Chen Gao and Yihong Chen and Depeng Jin and Yong Li},
booktitle={ICLR},
year={2021},
}
\balance
\appendix


\section{Overlapping Community Detection}
\label{appendix-slpa}

Algorithm~\ref{alg:slpa} outlines our overlapping community detection process. In brief, we first detect structural non-overlapping communities in the social network using the Leiden algorithm \cite{Traag_2019}, a scalable and effective method for non-overlapping community detection. To prevent unassigned users during this process, we ensure that all users belong to at least one community by assigning a self-node community to those who would otherwise remain unassigned. Finally, we convert the resulting non-overlapping user-community affiliation matrix into an overlapping version by applying the modularity-based conversion approach proposed in \cite{hieu2024overlappingcommunitydetectionalgorithms}. During this step, an additional hyperparameter $\theta$ is introduced to determine whether a specific user should be included in a new community. To obtain robust overlapping results, we set $\theta = 1.5$.


\section{Time Complexity Analysis}
\label{appendix:timecomplexity}
The main computational cost of \method can be divided into three parts: (1) pre-calculation for community detection, (2) forward propagation, and (3) the self-supervised loss.


For generating the user-community matrix $\mathbf{G}$, we adopt a two-step algorithm with complexity $O(|\mathcal{E}_S|\log(m) + |\mathcal{C}|m)$, 
where $|\mathcal{E}_S|$, $|\mathcal{C}|$, and $m$ denote the numbers of edges in the
social network, communities, and users, respectively.
Empirically, this step takes about 4 seconds on Douban-Book, 13 seconds on Yelp, and 9 seconds on Epinions, which is negligible compared to the total training time.

During training, \method generates user and item representations to compute the main recommendation loss. This step has a time complexity of $O(((L+1)|\mathcal{E}_R|+|\mathcal{E}_G| +|\mathcal{E}_S|)d +md^2)$, where $|\mathcal{E}_R|$, $|\mathcal{E}_S|$, and $|\mathcal{E}_G|$ are the numbers of edges in the user-item interaction graph, the social network, and the user-community affiliation graph, respectively, and $d$ is the embedding dimension. Most of the computational costs arise from LightGCN propagation ($O(L|\mathcal{E}_R|d)$). This forward operation is also required to generate contrastive views for the self-supervised objective, which involves masking random edges in $\mathbf{G}$, introducing only negligible additional overhead.


Finally, computing the InfoNCE loss for self-supervised learning requires processing batch representations and computing similarity scores, with complexity $O(Bd + Bmd)$, where $B$ is the batch size, $m$ is the number of users, and $d$ is the embedding dimension.
\section{Experimental Details}

\label{appendix-exp-details}

To ensure fair comparison, we adopt a unified evaluation pipeline with shared data splits and evaluation metrics, modifying baseline settings only when strictly necessary. However, this framework differs from the original experimental setups of several baselines reported in the official source code, which may lead to variations in the results. The specific variants are described below.
\begin{itemize}[left=0cm]
\item \textbf{DSL}: The official implementation uses test data for early stopping. 
We modified it to use validation data.
\item \textbf{RecDiff}: The official implementation uses test data for early stopping. We modified it to use validation data.
\item \textbf{GDMSR}: The official implementation performs negative sampling using the entire dataset, including both validation and test data, while we use only the training data.
The original code also uses leave-one-out splitting \cite{zhao2020revisiting} to define the test set, which is an easier setting than our 6:2:2 split.
\end{itemize}

\begin{algorithm}[h]
\caption{Two-Step Overlapping Community Detection}
\label{alg:slpa}
\begin{algorithmic}[1]
    \STATE \textbf{Input:} $\mathcal{G}_S=(\mathcal{V}_S,\mathcal{E}_S)$, $\mathcal{D}=\{d_S(u)|u\in \mathcal{V}_S\}$, Threshold $\theta$
    \STATE \textbf{Output:} User-Community Affiliation Matrix $\mathbf{G}$
    \STATE
    \STATE \textbf{Step 1. Get Communities by Leiden Algorithm}
    \STATE $\mathbf{G} \in \{0,1\}^{|\mathcal{V}_S|\times|\mathcal{C}|}$, $\mathcal{C}$ $\gets$ 
    $\mathbf{Leiden}(\mathcal{G}_S)$ 
    \COMMENT{Apply the Leiden algorithm to get the non-overlapping communities}

    \STATE
    \STATE \textbf{Step 2. Convert to Overlapping Communities}
    \STATE Let $d=\sum_{u\in \mathcal{V}_S}d_S(u)$.
    \STATE Let community of user $v$ as $c_v$.
    \STATE Iter $\gets$ True
    \STATE \textbf{while} Iter \textbf{do}
    \STATE \quad Iter $\gets$ False
    \STATE \quad \textbf{for each} $u\in\mathcal{V}_S$ \textbf{do}
    \STATE \quad \quad \textbf{for each} $c_j \in \{c_v|v\in \mathcal{N}_S(u),c_v\neq c_u\}$
    \STATE \quad \quad \quad $\mathbf{LHS}\gets \sum_{w\in c_j}\frac{\mathbf{1}((u,w)\in\mathcal{E}_S)}{d_S(u)}$
\STATE \quad \quad \quad $\mathbf{RHS}\gets \theta\sum_{w\in c_j}\frac{d_S(w)}{d}$
\STATE \quad \quad \quad \textbf{if} $\mathbf{LHS} > \mathbf{RHS}$ \textbf{then}
\STATE \quad \quad \quad \quad Add user $u$ to community $c_j$
\STATE \quad \quad \quad \quad $g_{u,c_j}\gets1$
\STATE \quad \quad \quad \quad Iter $\gets$ True
\STATE \quad \quad \quad \textbf{end if}
\STATE \quad \quad \textbf{end for}
\STATE \quad \textbf{end for}
\STATE \textbf{end while}

\STATE \textbf{Return:} $\mathbf{G}$
\end{algorithmic}
\end{algorithm}
\clearpage

\section{Additional Experiments}
\label{appendix:ablation}

\subsection{Ablation Study on Components of \method}

To evaluate the effectiveness of each component in \method, we conduct an ablation
study by removing or replacing individual modules:
\ding{172} removes the \moduleindividual module;
\ding{173} replaces the community-based representation with a LightGCN-based one and
adopts a SimGCL-style self-supervised objective~\cite{yu2022graph};
\ding{174} replaces the adaptive fusion with simple summation;
\ding{175} replaces the gating mechanism with an MLP;
and \ding{176} removes the SSL regularization.
Results on all datasets are reported in Table~\ref{table-ablation-module}.

\begin{table}[t]
\small
\setlength{\tabcolsep}{4pt} 
\renewcommand{\arraystretch}{0.9} 
\begin{center}
\caption{
Ablation study on removing or replacing components in \method.
\method achieves the lowest average rank, highlighting the benefit of all components.
}
\vspace{-10pt}
\label{table-ablation-module}
\resizebox{\columnwidth}{!}{
\begin{tabular}{l|cc|cc|cc|c}
\toprule
\multirow{2}{*}{Method} &
\multicolumn{2}{c|}{Douban-Book} &
\multicolumn{2}{c|}{Yelp} &
\multicolumn{2}{c|}{Epinions} &
\multirow{2}{*}{\makecell{Avg.\\Rank}} \\
\cmidrule{2-7}
& R@20 & N@20 & R@20 & N@20 & R@20 & N@20 & \\
\midrule
\textbf{\method (ours)}
& .1562 & \underline{.1330}
& \underline{.1214} & \underline{.0706}
& \underline{.0727} & \underline{.0421}
& \textbf{2.33} \\

\midrule
\ding{172}
& \textbf{.1590} & \textbf{.1344}
& .1167 & .0682
& .0668 & .0384
& 3.67 \\

\ding{173}
& .1557 & .1317
& .1211 & \textbf{.0709}
& \textbf{.0748} & \textbf{.0427}
& \underline{2.67} \\

\ding{174}
& .1567 & \underline{.1330}
& \textbf{.1216} & .0704
& .0692 & .0401
& 2.83 \\

\ding{175}
& \underline{.1568} & .1328
& .1155 & .0670
& .0697 & .0402
& 3.33 \\

\ding{176}
& .0738 & .0589
& .0908 & .0505
& .0499 & .0284
& 6.00 \\

\bottomrule
\end{tabular}
}
\end{center}
\vspace{-5pt}
\end{table}

\begin{table}[t]
\small
\setlength{\tabcolsep}{4pt}
\renewcommand{\arraystretch}{0.9}
\begin{center}
\caption{
Performance under adversarial social-noise settings.
\method consistently outperforms baselines across datasets,
demonstrating robustness to adversarial corruption.
}
\vspace{-10pt}
\label{table-adversarial-noise}
\resizebox{\columnwidth}{!}{
\begin{tabular}{l|cc|cc|cc}
\toprule
\multirow{2}{*}{Method} &
\multicolumn{2}{c|}{Douban-Book} &
\multicolumn{2}{c|}{Yelp} &
\multicolumn{2}{c}{Epinions} \\
\cmidrule{2-7}
& R@20 & N@20 & R@20 & N@20 & R@20 & N@20 \\
\midrule
MHCN 
& \underline{.1402} & .1159 
& .1056 & .0617 
& .0660 & .0381 \\
DSL  
& .1217 & .0993 
& .0952 & .0537 
& .0600 & .0339 \\
GBSR 
& .1355 & \underline{.1170}
& \underline{.1099} & \underline{.0622}
& \underline{.0679} & \underline{.0391} \\
\midrule
\textbf{\method (ours)}
& \textbf{.1573} & \textbf{.1335}
& \textbf{.1209} & \textbf{.0710}
& \textbf{.0700} & \textbf{.0410} \\
\bottomrule
\end{tabular}
}
\end{center}
\vspace{-5pt}
\end{table}


We observe that \method with all components achieves the most robust performance across all datasets. For \ding{172}, excluding \moduleindividual yields better performance on Douban-Book but leads to performance drops on Yelp and Epinions. Conversely, for \ding{173}, removing \modulecommunity reduces performance on Douban-Book but improves it on Epinions. 
We attribute these differences to dataset characteristics. Yelp and Epinions have relatively large social networks, where behavior embeddings between users are more reliable, allowing socially connected item information to provide robust and diverse signals to the target user. In contrast, Douban-Book has a smaller social network, where community-level signals are more effective than individual item signals. 
For \ding{174} and \ding{175}, replacing the adaptive gating mechanism with simple summation or an MLP results in inconsistent performance across datasets. In particular, on the Epinions dataset, which exhibits denser social relations, dynamic fusion is more effective than static summation. Moreover, although the MLP-based fusion is more expressive, the lack of explicit regularization leads to unstable representations and overall performance degradation, highlighting the importance of properly regularized fusion mechanisms. These results indicate that adaptively fusing heterogeneous social information is
critical to the effectiveness of our method.
Finally, for \ding{176}, removing the SSL regularization component causes the most severe performance degradation across all datasets, demonstrating that self-supervised regularization is essential for stabilizing community signals and
maintaining consistent and robust user representations.


\begin{figure}[t]
\centering

\includegraphics[width=0.95\columnwidth]{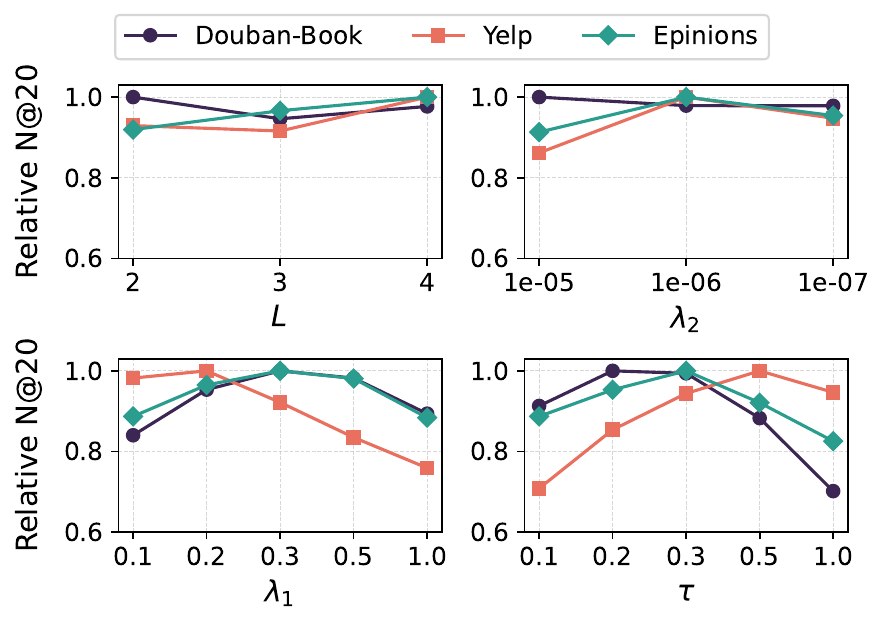}
\vspace{-15pt}
\caption{Hyperparameter analysis of \method. The model shows robustness with respect to the number of layers $L$ and the regularization weight $\lambda_{2}$, while requiring dataset-specific tuning for the SSL weight $\lambda_{1}$ and the temperature $\tau$.}
\label{fig-hyperparameter-sensitivity}
\Description[Hyperparameter sensitivity of PULSE across datasets]{
Four line charts showing the sensitivity of recommendation performance to different hyperparameters across three datasets: Douban-Book, Yelp, and Epinions.
In all charts, the vertical axis represents relative normalized discounted cumulative gain at rank twenty.
The top-left chart varies the number of layers from two to four, showing stable performance across all datasets.
The top-right chart varies the regularization weight lambda two from one times ten to the minus five to one times ten to the minus seven, with little change in relative performance.
The bottom-left chart varies the self-supervised learning weight lambda one from zero point one to one point zero, where performance peaks at intermediate values and declines at higher values depending on the dataset.
The bottom-right chart varies the temperature parameter tau from zero point one to one point zero, showing dataset-dependent trends with optimal performance at intermediate values.
Overall, the figure shows that performance is robust to changes in the number of layers and the regularization weight, but more sensitive to the self-supervised learning weight and the temperature parameter.
}
\vspace{-5pt}
\end{figure}

\subsection{Hyperparameter Sensitivity}
\label{ssec:hyperparameer-sensitivity}
To investigate the sensitivity of \method to its key hyperparameters, we conduct experiments by varying four parameters: the number of layer $L$, the regularization weight $\lambda_2$, the SSL weight $\lambda_1$, and the temperature $\tau$ in the SSL objective.
For each setting, we report the relative N@20, normalized by the best-performing configuration. 

Figure \ref{fig-hyperparameter-sensitivity} presents the results. The SSL-related parameters, $\lambda_1$ and $\tau$, exhibit clear dataset-dependent sensitivity. On Douban-Book and Epinions, \method achieves the highest relative N@20 when $\lambda_1=0.3$, whereas the optimal performance on Yelp is obtained with $\lambda_1=0.2$. A similar trend is observed for the temperature parameter: the best results occur at $\tau=0.2$ for Douban-Book, $\tau=0.3$ for Epinions, and $\tau=0.4$ for Yelp. 
This suggests that dataset-specific tuning is required to fully leverage the benefits of contrastive learning.

In contrast, the structural and regularization-related hyperparameters demonstrate stronger robustness across datasets. \method maintains stable performance across a wide range of layer depths, showing that its effectiveness is not heavily dependent on fine-grained adjustments to $L$. Likewise, the regularization weight $\lambda_2$ exhibits consistent results regardless of the dataset. These observations indicate that while \method requires careful tuning of SSL-related parameters for optimal performance, it remains inherently stable with respect to architectural depth and regularization strength.

\subsection{Robustness to Adversarial Social Noise}
\label{appendix:adversarial-noise}
To further examine the robustness of \method in social noise in Section~\ref{ssec:robustness-to-social-noise}, we evaluate \method under adversarial social-noise settings. This adversarial setup is specifically designed to deliberately disrupt community structures in social networks~\cite{liu2019rem}, rather than introducing random perturbations.

As shown in Table~\ref{table-adversarial-noise}, \method consistently outperforms
SocialRec baselines across all three datasets, highlighting the advantage of
\method's design. On average, \method achieves an 11.0\% improvement in NDCG@20
over the strongest competitor.
This robustness can be attributed to two key components: (i) the adaptive integration of diverse social signals via a gating network, and (ii) community-aware self-supervised learning. Even when community structures are perturbed, the proposed self-supervised objective enables meaningful user-related signals to be preserved in user representations, while the gating network effectively regulates the influence of noisy signals.



\subsection{Comparison with  Compression-based Methods}
\label{appendix:compression}
To demonstrate that the benefits of \method extend beyond parameter efficiency in graph collaborative filtering, we compare \method with general-purpose model compression techniques applied to LightGCN:
\ding{172} embedding dimension reduction to match the number of parameters of \method,
\ding{173} Low-Rank Adaptation (LoRA)~\cite{hu2022lora}, and
\ding{174} PEP~\cite{liu2021learnable}.
These methods aim to reduce parameter storage without modifying the underlying
user modeling paradigm.

As shown in Table~\ref{table-compression-main}, compression-based baselines suffer
substantial performance degradation across all datasets.
The degradation becomes even more pronounced in the cold-start setting
(Table~\ref{table-coldstart-compression}).
These results indicate that naively compressing existing user embeddings is
insufficient for effective recommendation, especially when user interaction data
are sparse.
In contrast, \method achieves parameter efficiency by rethinking how user representations are constructed through socially derived signals, including community-level information, rather than compressing trainable user embeddings.
By generating user representations via a lightweight and socially informed
mechanism, \method maintains strong recommendation performance while dramatically reducing the number of parameters required for user modeling.

\begin{table}[t]
\small
\setlength{\tabcolsep}{4pt}
\renewcommand{\arraystretch}{0.9}
\begin{center}
\caption{
Performance comparison with general-purpose compression methods.
\method consistently outperforms compression-based variants across all datasets.}
\vspace{-10pt}
\label{table-compression-main}
\resizebox{\columnwidth}{!}{
\begin{tabular}{l|cc|cc|cc}
\toprule
\multirow{2}{*}{Method} &
\multicolumn{2}{c|}{Douban-Book} &
\multicolumn{2}{c|}{Yelp} &
\multicolumn{2}{c}{Epinions} \\
\cmidrule{2-7}
& R@20 & N@20 & R@20 & N@20 & R@20 & N@20 \\
\midrule
\ding{172}
& \underline{.1218} & \underline{.0975}
& \underline{.0950} & \underline{.0532}
& \underline{.0595} & \underline{.0339} \\

\ding{173}
& .0852 & .0680
& .0761 & .0414
& .0435 & .0238 \\

\ding{174}
& .1095 & .0855
& .0890 & .0497
& .0562 & .0313\\

\midrule
\textbf{\method (ours)}
& \textbf{.1562} & \textbf{.1330}
& \textbf{.1214} & \textbf{.0706}
& \textbf{.0727} & \textbf{.0421} \\
\bottomrule
\end{tabular}
}
\end{center}
\vspace{-5pt}
\end{table}
\begin{table}[t]
\huge
\begin{center}
\caption{
Cold-start performance comparison between \method and compression-based baselines.
\method consistently outperforms compressed  variants in cold-start settings.}
\vspace{-10pt}
\label{table-coldstart-compression}
\resizebox{\columnwidth}{!}{
\begin{tabular}{l|cccc|cccc}
\toprule
\multirow{2}{*}{Method} &
\multicolumn{4}{c|}{Douban-Book} &
\multicolumn{4}{c}{Yelp} \\
\cmidrule{2-9}
& R@10~ & R@20~ & N@10~ & N@20~
& R@10~ & R@20~ & N@10~ & N@20~ \\
\midrule
\ding{172}
& .0191~ & .0371~ & .0202~ & .0259~
& .0205~ & .0267~ & .0128~ & .0150~ \\

\ding{173}
& .0203~ & .0296~ & .0231~ & .0254~
& \underline{.0212}~ & \underline{.0330}~ & \underline{.0142}~ & \underline{.0180}~ \\

\ding{174}
& \underline{.0345}~ & \underline{.0526}~ & \underline{.0354}~ & \underline{.0397}~
& .0008~ & .0016~ & .0016~ & .0018~ \\
\midrule
\textbf{\method (ours)}
& \textbf{.0387}~ & \textbf{.0617}~ & \textbf{.0499}~ & \textbf{.0542}~
& \textbf{.0666}~ & \textbf{.0922}~ & \textbf{.0426}~ & \textbf{.0507}~ \\
\bottomrule
\end{tabular}
}
\end{center}
\vspace{-5pt}
\end{table}

\newpage



\end{document}